\documentclass[twocolumn]{emulateapj}

\usepackage[bookmarks=true]{hyperref}
\usepackage{natbib} 

\begin{document}

\renewcommand{\arraystretch}{1.2}

\newcommand{\QSOKim}{K-method} 

\newcommand{\NCanAll}{2,566} 
\newcommand{\NCanSAGE}{469} 
\newcommand{\NCanPhotoZStar}{1274} 
\newcommand{\NCanPhotoZGal}{686} 
\newcommand{\NCanPhotoZ}{602} 
\newcommand{\NCanPhotoZGalFit}{84} 
\newcommand{\NCanXray}{60} 
\newcommand{\NCanSum}{393} 
\newcommand{\NCanStrong}{663}
\newcommand{\increasedRatio}{12}

\title{A Refined QSO Selection Method Using Diagnostics Tests:\\ 
\NCanStrong{}  QSO Candidates in the LMC}

\author{Dae-Won Kim\altaffilmark{1}\altaffilmark{,2}\altaffilmark{,3},
Pavlos Protopapas\altaffilmark{1}\altaffilmark{,3},
Markos Trichas\altaffilmark{1},
Michael Rowan-Robinson\altaffilmark{4},
Roni Khardon\altaffilmark{5},
Charles Alcock\altaffilmark{1},
Yong-Ik Byun\altaffilmark{2}\altaffilmark{,6}}
\affil{\altaffilmark{1}Harvard-Smithsonian Center for Astrophysics, Cambridge, MA, USA}
\affil{\altaffilmark{2}Department of Astronomy, Yonsei University, Seoul, South Korea}
\affil{\altaffilmark{3}Institute for Applied Computational Science, Harvard University, Cambridge, MA, USA}
\affil{\altaffilmark{4}Astrophysics Group, Imperial College, London, United Kingdom}
\affil{\altaffilmark{5}Department of Computer Science, Tufts University, Medford, MA, USA}
\affil{\altaffilmark{6}Yonsei University Observatory, Yonsei University, Seoul, South Korea}

\begin{abstract}

We present \NCanStrong{} QSO candidates in the Large Magellanic Cloud (LMC)
selected using multiple diagnostics.
We started with a set of \NCanAll{} QSO candidates selected
using the methodology presented in our previous work based on time variability of the MACHO LMC lightcurves.
We then obtained additional information
for the candidates by crossmatching them with the Spitzer SAGE,
the 2MASS, the Chandra, the XMM, and an LMC UBVI catalog.
Using this information, we specified six diagnostic features based on mid-IR colors,
photometric redshifts using SED template fitting, and X-ray luminosities
in order to further discriminate high confidence QSO candidates in the absence of spectra information.
We then trained a one-class SVM (Support Vector Machine) model
using the diagnostics features of the confirmed 58 MACHO QSOs.
We applied the trained model to the original candidates and
finally selected \NCanStrong{} high confidence QSO candidates.
Furthermore, we crossmatched these \NCanStrong{} QSO candidates with the newly confirmed 
151 QSOs and 275 non-QSOs in the LMC fields.
On the basis of the counterpart analysis, 
we found that the false positive rate is less than 1\%.

\end{abstract}

\keywords{Magellanic Clouds - methods: data analysis - quasars: general}

\section{Introduction}

Active Galactic Nuclei (AGNs) 
are very energetic extragalactic objects 
that have been studied in many astronomical fields
such as galaxy formation and evolution  
(e.g. \citealt{Heckman2004ApJ, Bower2006MNRAS, Trichas2009MNRAS, Trichas2010MNRAS}),
large scale structure (e.g. \citealt{Ross2009ApJ}), 
dark matter substructure (e.g. \citealt{Miranda2007MNRAS}),
and black hole growth (e.g. \citealt{Kollmeier2006ApJ}).
%Thus having a large set of QSOs is crucial for studying
%observational cosmology.

It is known that QSOs show strong variability over
wide range of wavelengths on a time scale 
from a few days to several years \citep{Hook1994MNRAS, Hawkins2002MNRAS}.
It is widely believed that the variability is associated with
accretion disk instability \citep{Rees1984ARAA, Kawaguchi1998ApJ}.
Recently, interesting studies on QSO variability
have been published \citep{Kelly2009ApJ, MacLeod2010ApJ},
which confirmed a correlation between the time scale of QSO variability and the physical parameters of 
QSOs such as black hole mass.
Although these studies confirmed the correlation, 
different studies showed a discrepancy at the time scales of QSO variability 
\citep{Kelly2009ApJ, Kozlowski2010ApJ, MacLeod2010ApJ}. 
Possible reasons for the discrepancy are
1) poorly-sampled lightcurves and/or short observational periods, 
2) false positives such as stellar contaminations in their QSO candidates, 
and 3) biased QSO samples in luminosity or black hole mass.
Thus having a well-sampled set of QSO  lightcurves with  a long baseline
and small number of  false positives is critical 
for the comprehensive analysis of this correlation.
Note that there are only a few hundreds well-sampled QSO lightcurves, 
and a large portion of them are around the LMC fields 
where the MACHO  survey monitored for several years
(e.g. see \citealt{Geha2003AJ, Kelly2009ApJ, Kozlowski2011arXiv}).

The MACHO survey observed the sky around the LMC for 7.4 years
with relatively regular sampling of a few days. The majority of the MACHO
lightcurves have more than several hundred data points and 
therefore the MACHO lightcurves are suitable for the QSO variability studies.
Nevertheless, there are only 59 confirmed MACHO QSOs
in the 40 deg$^{2}$ areas around the LMC \citep{Geha2003AJ}.
The main reasons for the relatively small number 
of QSOs are 1) the crowdedness of the fields,
which makes it difficult to select QSO candidates among the dense stellar sources
and thus yields a high false positive rate (e.g. see \citealt{Geha2003AJ, Dobrzycki2005AA}),
and 2) the high cost of spectroscopic or X-ray observations, 
which are the best methods for confirming QSOs.
Thus a novel QSO selection algorithm with a high efficiency
and a low false positive rate is essential to 
make the best use of the expensive spectroscopic telescope time
and increase the collection of QSOs.

In our previous work \citep{Kim2011ApJ}, we developed a QSO selection method
using a supervised classification model trained on a set
of variability features extracted from the MACHO lightcurves
including a variety of variable stars, non-variable stars and QSOs.
The trained model showed  high efficiency of 80\% and  low false positive rate of 25\%.
Using this method, we first selected \NCanAll{} QSO candidates from the lightcurve database.
We then developed and employed a decision procedure on the basis of diagnostics 
using 1) mid-IR colors, 2) photometric redshifts, and 3) X-ray luminosities on these candidates
in order to separate {\em{high confidence}} QSO candidates
(hereinafter hc-QSOs).
As a result, we chose in total \NCanStrong{} hc-QSOs out of \NCanAll{}.
These \NCanStrong{} candidates are likely QSOs;
if confirmed this will increase the previous collection of QSOs in
the MACHO LMC database by a factor of  $\sim$\increasedRatio{}.
Note that most of the hc-QSO lightcurves are well-sampled for 7.4 years
(i.e. several hundreds data points with relatively regular sampling).
and are   chosen in such 
a way to exclude any potential false positives. 
Therefore the lightcurve collection of hc-QSOs is a valuable set for QSO variability studies
and can be used as a target set for spectroscopic observations.

In Section \ref{sec:MACHO}, we briefly introduce 
the MACHO database and the QSO selection algorithm 
that we developed to select the initial set of QSO candidates.
We then present multiple diagnostics that we applied on
the set of QSO candidates in Section \ref{sec:diagnostics}.
Section \ref{sec:finalCatalog} presents
a classification model trained on
the diagnostics features in order to choose hc-QSOs.
In Section \ref{sec:koz_QSOs},
we crossmatch our candidates with newly discovered
QSOs in the LMC fields.
A summary is given in Section \ref{sec:summary}.

\section{QSO Candidates in the MACHO LMC database}
\label{sec:MACHO}

We first selected QSO candidates from the MACHO lightcurve database
using the QSO selection method developed by \citet{Kim2011ApJ} (hereinafter, \QSOKim{}).
In this paper we used a 10\% QSO probability product
cut to select the QSO candidates rather than a 25\% cut which \citet{Kim2011ApJ} used because we will employ  other diagnostics (see Section \ref{sec:diagnostics}) that 
are able to effectively remove  false positives.\footnote{
A lower probability cut typically produces not only more QSO candidates
but also more false positives.} 
Here probability product is the  product of the probabilities 
derived  independently from MACHO B and R band lightcurves
using  Support Vector Machine \citep{Boser1992} and  Platt's probability estimation \citep{Platt1999}.
By definition, QSO candidates with higher probabilities are more likely to be QSOs.
With the probability cut of 10\%, we found \NCanAll{} QSO candidates.

\section{Diagnostics of the QSO Candidates}
\label{sec:diagnostics}

In the following subsections, we will introduce the diagnostics  performed and the consequent results.

\subsection{Spitzer mid-IR Properties}
\label{sec:mid-IR}

It is known that mid-IR color selection is an efficient discriminator for 
AGNs and stars/galaxies resulting from the fact that  the spectral energy distributions of these sources
are substantially different from each other \citep{Laurent2000AA, Lacy2004ApJS}.
\citet{Lacy2004ApJS}
introduced a mid-IR color cut to separate AGNs using Spitzer SAGE
(Surveying the Agents of a Galaxy's Evolution; \citealt{Meixner2006AJ}) catalog.
\citet{Kozlowski2009ApJ}  employed a similar mid-IR color cut
and selected about 5,000 AGN candidates from the Spitzer SAGE catalog.

We used these mid-IR color selections as the first diagnostic. We 
crossmatched our candidates with the Spitzer SAGE LMC catalog
containing 6 million mid-IR objects in order to check 
whether our candidates are inside the mid-IR selection cuts.
We searched for the nearest SAGE source from each candidate 
within an 1$^{\prime\prime}$ search radius.
In order to minimize false crossmatchings, we defined a source as a counterpart 
only if there are no other Spitzer sources within a 3$^{\prime\prime}$ radius from the candidate.

\begin{figure*}
\begin{center}
\begin{minipage}[c]{8cm}
        \includegraphics[width=1.0\textwidth]{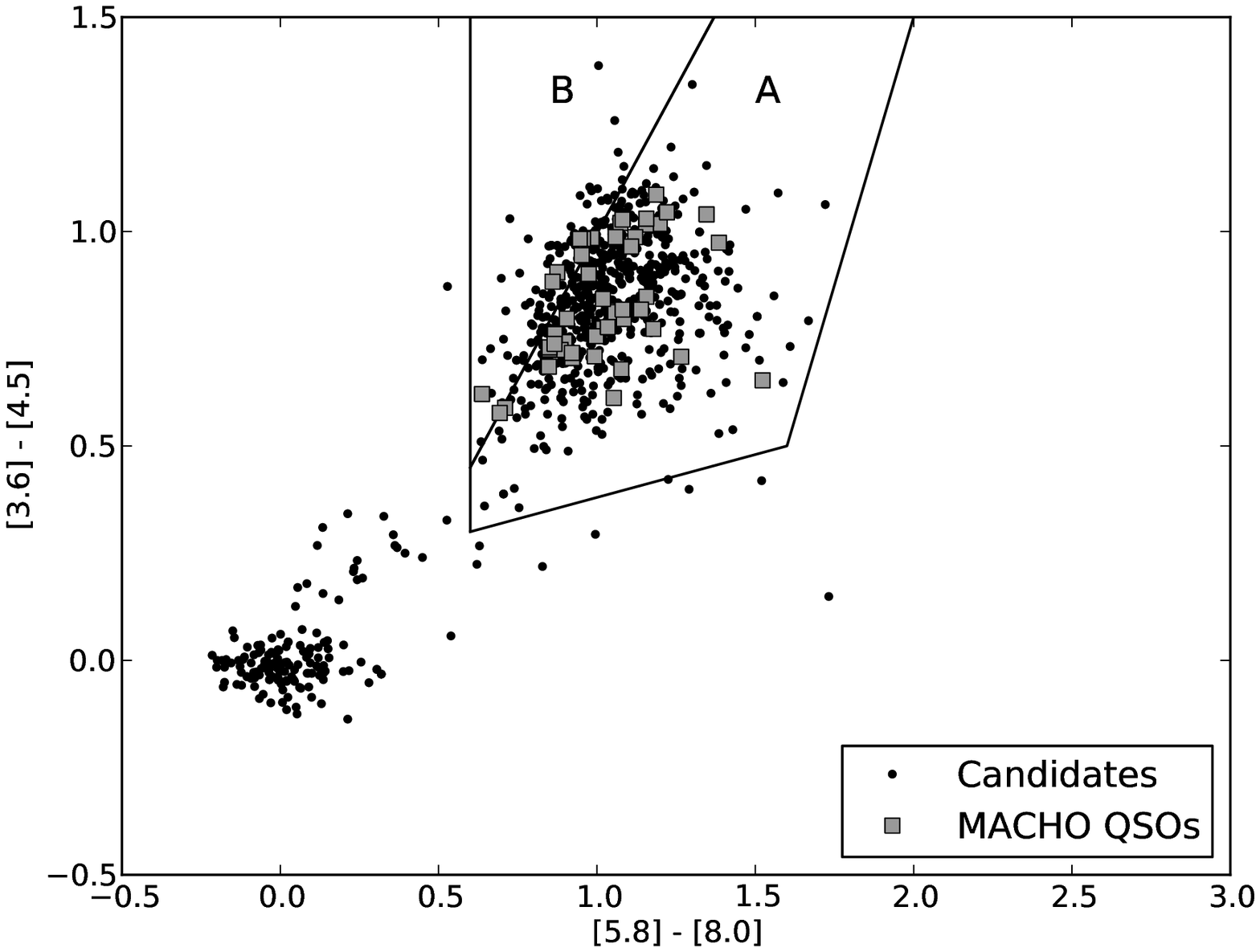} %./MACHO/ML/plot_sage_candidates_with_macho.py
\end{minipage}
\begin{minipage}[c]{8cm}
        \includegraphics[width=1.0\textwidth]{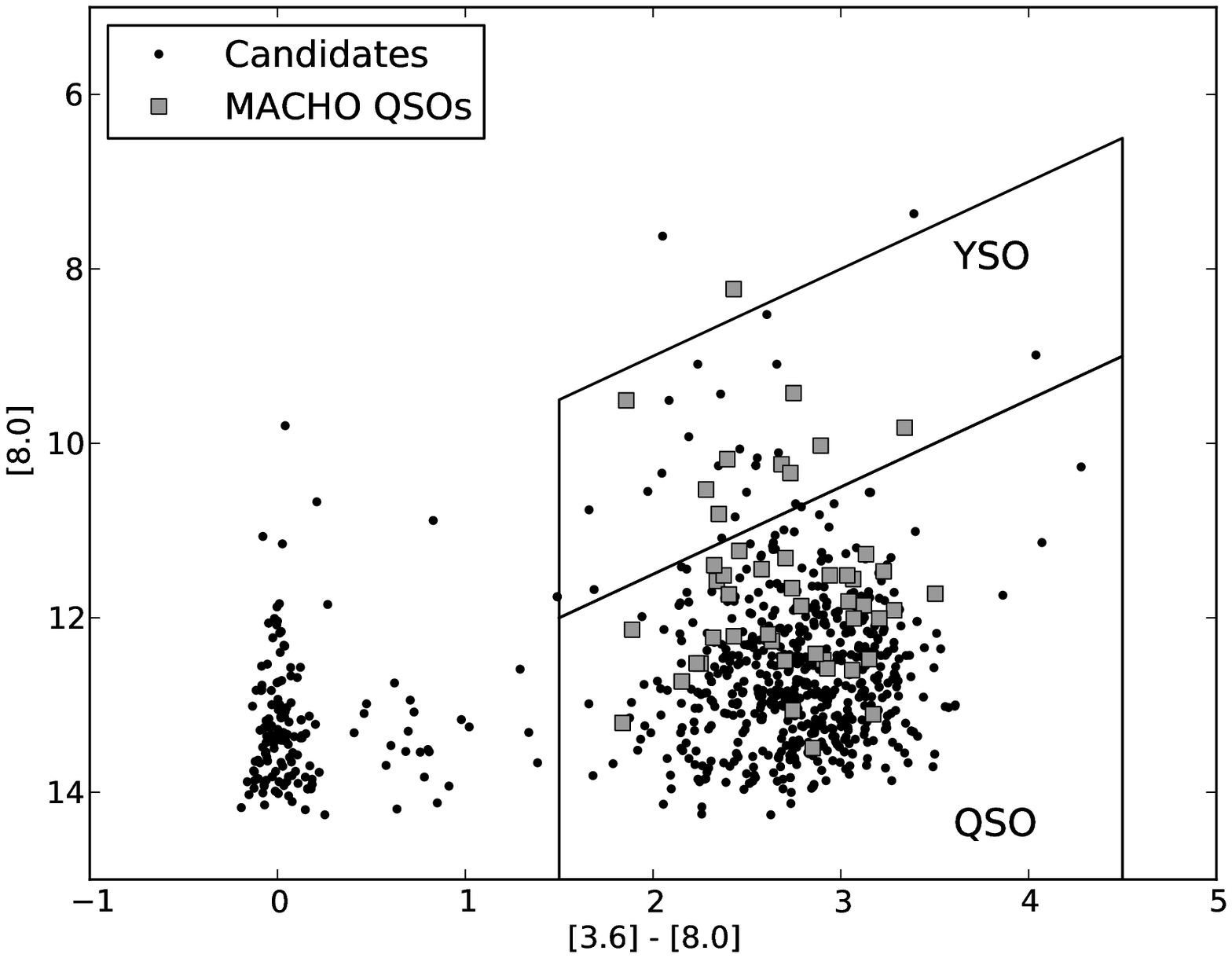}
\end{minipage}
\end{center}
    \caption
           {Mid-IR color-color and color-magnitude diagrams of the Spitzer SAGE counterparts 
           with our QSO candidates (dots).
           Each axis of the figure is either Spitzer magnitude or color.
           All sources inside the four regions A, B, QSO and YSO are potential QSOs \citep{Kozlowski2009ApJ}.
           There are \NCanSAGE{} candidates inside the both QSO and A regions,
           which are the most promising QSO candidates.
           The confirmed MACHO QSOs are also inside these four regions (boxes).}
    \label{fig:sage_can_CCD_CMD}
\end{figure*}

We found about 700 Spitzer counterparts
shown in Figure \ref{fig:sage_can_CCD_CMD} (dots).
The sources inside region B could
either be  AGNs or  stars, 
while the sources inside region A are likely AGNs. 
The YSO region is thought to be dominated by  Young Stellar Objects (YSOs) while
the QSO region is thought to be dominated by AGNs.
Nevertheless, all the sources inside these four regions 
are potential QSOs.\footnote{The strongest statement is that
QSOs are very unlikely to be outside those four regions.}
Almost all of the confirmed MACHO QSOs are inside these four regions 
as shown in Figure \ref{fig:sage_can_CCD_CMD} (boxes).\footnote{There
are 48 MACHO QSOs that were crossmatched with the SAGE catalog.}
The candidates inside these regions are most likely 
broad emission line QSOs (i.e. Type I AGNs  \citep{Stern2005ApJ}).
Among these counterparts, the sources inside both the QSO and the A regions
are likely to be QSOs.
We found that \NCanSAGE{} QSO candidates are inside  both QSO and A regions.

\begin{figure}
\begin{center}
       \includegraphics[width=0.45\textwidth]{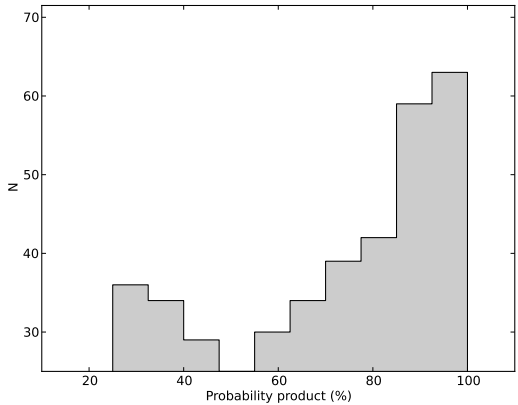} %./MACHO/ML/plot_sage_candidates.py
\end{center}
    \caption{Histogram of \QSOKim{} QSO probabilities of the SAGE counterparts inside both the QSO and the A 
    (see Figure \ref{fig:sage_can_CCD_CMD}). There are more high probability candidates
    than low probability candidates, which indicates that the candidates inside the QSO and the A
    are likely to be QSOs. The histogram also shows a bimodal distribution
    as  is addressed in Section \ref{sec:photometric_redshift}.}
    \label{fig:SAGE_prob}
\end{figure}

Figure \ref{fig:SAGE_prob} shows the estimated K-method QSO probability products
of these \NCanSAGE{} candidates. 
As the histogram shows, there are more QSO candidates 
at higher probability than lower probability, which
implies that the mid-IR diagnostic is  in line  with the \QSOKim{}.\footnote{In 
the case of the entire  \NCanAll{} QSO candidates, the number of candidates
decreases at higher probability.}
In addition, the histogram shows a bimodal distribution of the probabilities.
We will address this  bimodality  in the following section.

\subsection{Photometric Redshift Using Template Fitting}
\label{sec:photometric_redshift}

\begin{figure}
\begin{center}
       \includegraphics[width=0.45\textwidth]{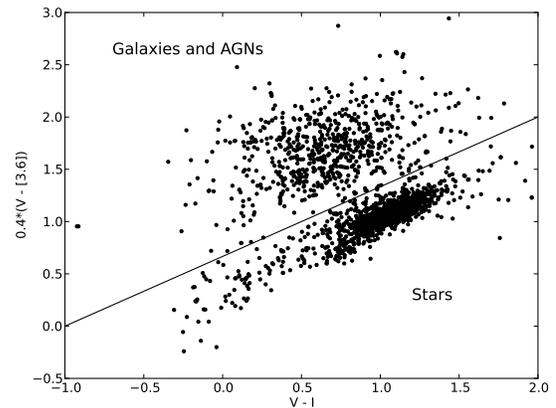} % ./MACHO/photoz/plot_photoz_all.py
\end{center}
    \caption{Criterion (the solid line) to separate extragalatic sources (`Galaxies and AGN' in the figure)
     from stars \citep{Eisenhardt2004ApJS, Rowan2005AJ}
    Using the criterion, 686 candidates were classified as extragalatic sources (above the line) and 
    1274 candidates were classified as stars (below the line).
     }
    \label{fig:RR_CCD}
\end{figure}

We first crossmatched the \NCanAll{}  QSO candidates 
with the UBVI catalog for the LMC \citep{Zaritsky2004AJ} 
and the 2MASS catalog \citep{Skrutskie2006AJ}
to extract UBVI and JHK magnitudes.
We searched the nearest source from each of 
the candidates within a 3$^{\prime\prime}$ search radius.
In the case of the UBVI catalog, 
we found in total 2,375 counterparts.
Among them, 84\% (93\%) UBVI counterparts are within 
a 1$^{\prime\prime}$ (1.5$^{\prime\prime}$)
distance from the candidates. In addition, 
only 0.3\% (2\% or 17\%) of the candidates have another
counterpart within an 1$^{\prime\prime}$ 
(1.5$^{\prime\prime}$ or 3$^{\prime\prime}$) distance from the candidates. 
Thus the portion of the false crossmatching is not significant.
In the case of the 2MASS catalog,
we found in total 846 counterparts.
From those, 74\% (83\%) are within a 1$^{\prime\prime}$ (1.5$^{\prime\prime}$)
distance from the candidates while 0\% (0.1\% or 0.5\%) of the candidates have another
counterpart within a 1$^{\prime\prime}$ 
(1.5$^{\prime\prime}$ or 3$^{\prime\prime}$) distance from the candidates. 
Again the portion of the false crossmatching is negligible.

We then separated stars from `Galaxies and AGNs' (i.e. extragalatic sources) using
a  criterion proposed by \citet{Eisenhardt2004ApJS} and \citet{Rowan2005AJ}.
Figure \ref{fig:RR_CCD} shows the criterion (the solid line) we applied.
There were \NCanPhotoZGal{} extragalatic sources (above the cut)
and \NCanPhotoZStar{}  stars (below the cut).\footnote{We
excluded the sources that do not have enough color information.}
These \NCanPhotoZGal{} extragalatic sources were then fitted with 
galaxy templates in order to derive photometric redshifts \citep{Rowan2008MNRAS}.
The templates contained three QSO, one starburst and 10 galaxy templates.
For details about the photometric redshift estimations and the SED  template fitting, 
see \citet{Rowan2008MNRAS}.

\begin{figure}
\begin{center}
       \includegraphics[width=0.45\textwidth]{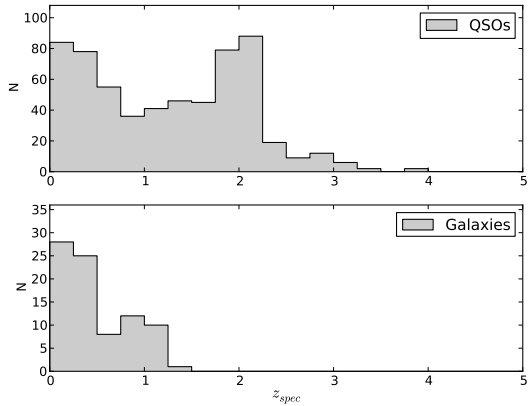} % ./MACHO/photoz/plot_photoz_all.py
\end{center}
    \caption{Photometric redshifts of the \NCanPhotoZ{} QSO candidates fitted with the AGN templates (the top panel)
    and the \NCanPhotoZGalFit{} QSO candidates fitted with the galaxy templates (the bottom panel) \citep{Rowan2008MNRAS}.
    The \NCanPhotoZ{} QSO candidates show relatively larger redshifts than the 84 candidates.
    }
    \label{fig:photz_hist}
\end{figure}

\begin{figure}
\begin{center}
       \includegraphics[width=0.45\textwidth]{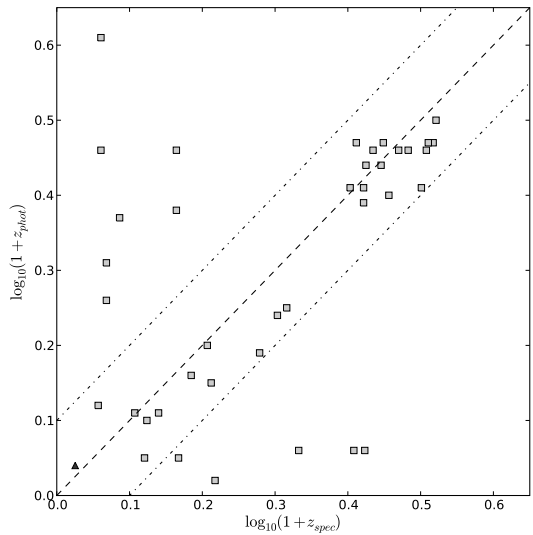} % ./MACHO/photoz/plot_photoz_all.py
\end{center}
    \caption{Comparison between the spectroscopic redshifts \citep{Geha2003AJ} 
    and the photometric redshifts for the confirmed MACHO QSOs.
    Seventy percent of estimated redshifts are well-matched with the spectroscopic
    redshifts (see the dashed line corresponding to  $\pm$0.1 dex accuracy).
    There is one MACHO QSO (triangle) that is fitted with the galaxy templates
    and 39 MACHO QSOs (squares) that are fitted with the AGN templates \citep{Rowan2008MNRAS}.
    }
    \label{fig:specz_photz}
\end{figure}

Among the extragalatic sources,
\NCanPhotoZ{} were fitted with AGN templates (i.e. QSOs)
while the remaining \NCanPhotoZGalFit{} were fitted with the galaxy templates (i.e. galaxies).
These \NCanPhotoZ{} candidates are likely QSOs.
Figure \ref{fig:photz_hist} shows the photometric
redshifts of these QSOs and galaxies. 
As the figure shows, the QSOs (the top panel) have relatively higher redshifts than the galaxies (the bottom panel).
QSOs are much more luminous than galaxies 
and thus are detectable at higher redshifts than galaxies.
In Figure \ref{fig:specz_photz}, we show
the comparison between the photometric redshifts
and the spectroscopic redshifts of the confirmed MACHO QSOs \citep{Geha2003AJ}.
Out of the 58 confirmed MACHO QSOs\footnote{Note that
58 of 59 MACHO QSOs had been monitored
more than several hundreds times during 7.4 years' observation
while the remaining one MACHO QSO has only about 50 data points. We
excluded the QSO with 50 data points from the analysis in this paper.}, 
40 are fitted with the
photometric redshift code. The remaining 18 were not 
fitted due to the lack of data (i.e. UBVI magnitudes).
Among these 40 confirmed MACHO QSOs, only one 
was best fitted with galaxy templates while the other 39 were
fitted with AGN templates. 
The QSO best fitted with the galaxy templates is
confirmed to be a  QSO from the works done by \citet{Schmidtke1999AJ, Geha2003AJ}.
Out of the 40 QSOs, 28 (70\%)
are inside the $\pm$0.1 dex accuracy (the dashed line in the figure).

\begin{figure*}
\begin{center}
\begin{minipage}[c]{8cm}
        \includegraphics[width=1.0\textwidth]{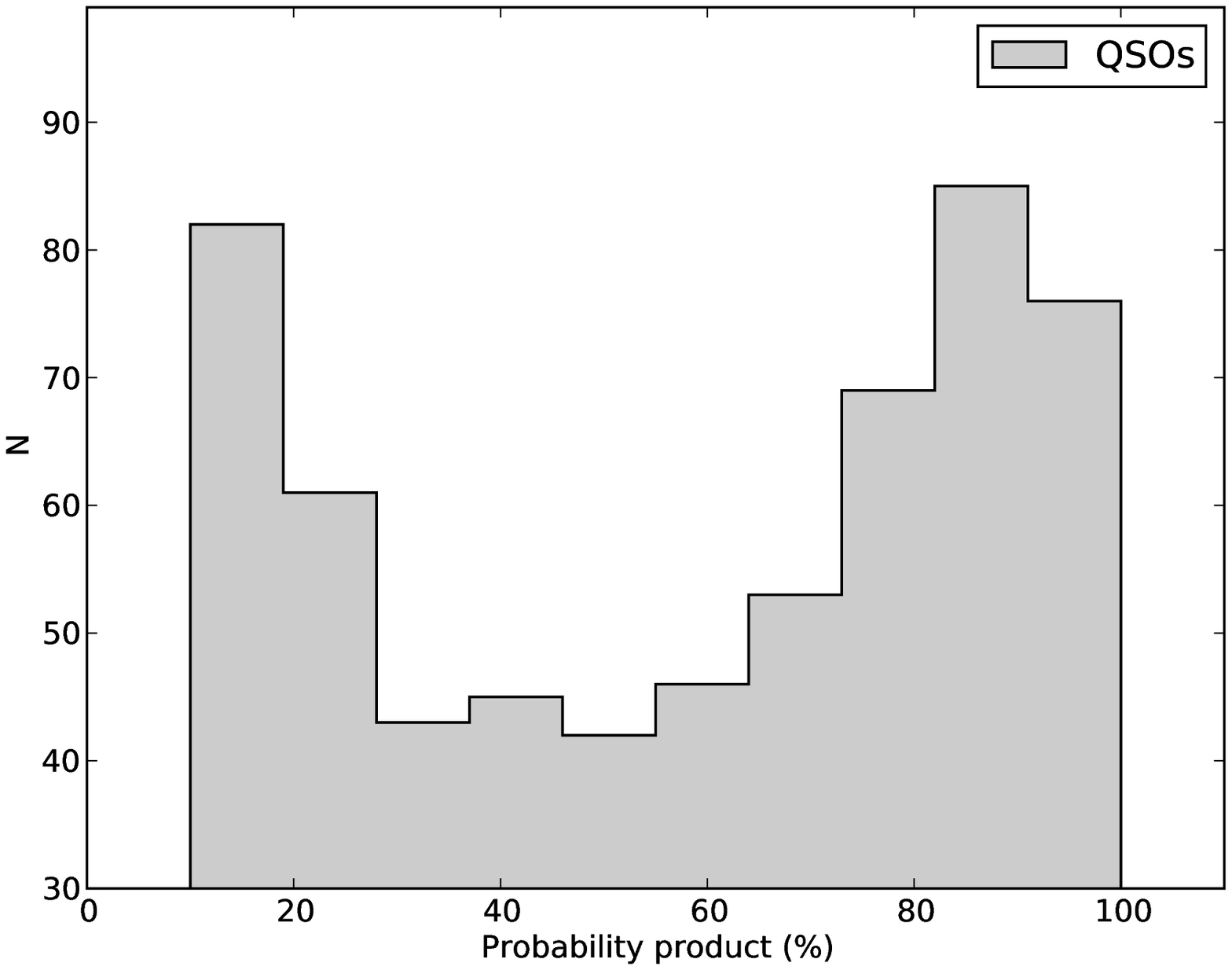} %./MACHO/photoz/plot_photoz_all.py
\end{minipage}
\begin{minipage}[c]{8cm}
        \includegraphics[width=1.0\textwidth]{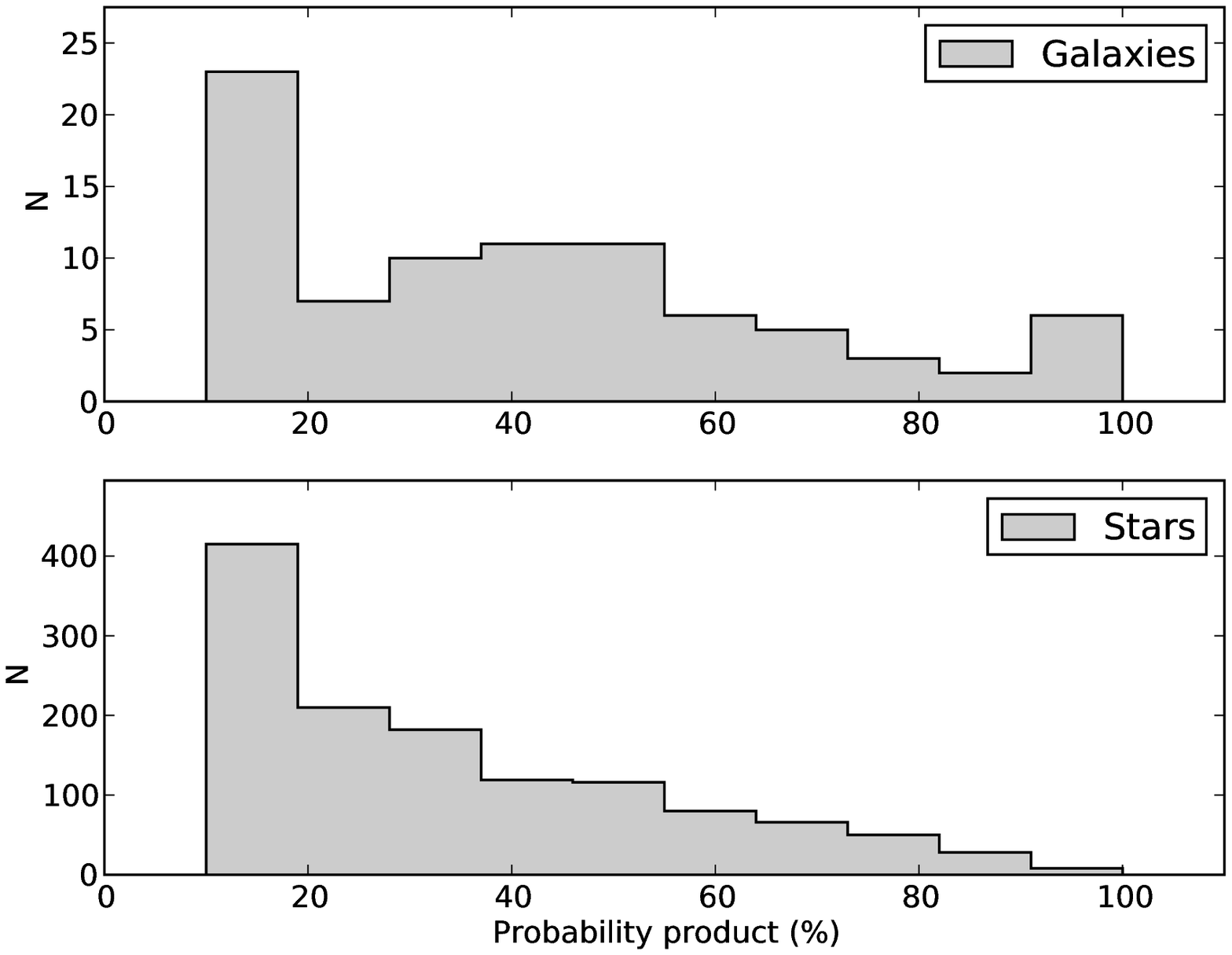}
\end{minipage}
\end{center}
    \caption
           {left: Histogram of the estimated  \QSOKim{} QSO probabilities for \NCanPhotoZ{} QSOs fitted with the AGN templates. 
           The histogram shows a bimodal distribution similar to the histogram shown in Figure \ref{fig:SAGE_prob}.
           The bimodality is correlated with different variability characteristics of the low and high probability QSO candidates.
           See the text and Figure \ref{fig:photoZ_feature_hist} for details.
           right: Histogram of the estimated \QSOKim{} QSO probability of \NCanPhotoZGalFit{} galaxies (the top panel)
           and \NCanPhotoZStar{} stars (the bottom panel)
           separated using a approach proposed by \citet{Eisenhardt2004ApJS} 
           and \citet{Rowan2005AJ}. As the histogram clearly shows, 
           they have relatively lower probabilities than QSOs.
           }
    \label{fig:photoZ_prob}
\end{figure*}

Figure \ref{fig:photoZ_prob} shows the \QSOKim{} probability of 
QSOs, galaxies and stars
discriminated during the photometric redshift estimation.
As the figure  shows, the majority of QSOs have higher probabilities than galaxies and stars,
which implies that galaxies and  stars have different  and most likely weaker variability characteristics 
from/than QSOs. Note that the probabilities are from  the \QSOKim{} which mainly used  variability features 
of lightcurves to select QSO candidates.

\begin{figure*}
\begin{center}
\begin{minipage}[c]{8cm}
        \includegraphics[width=1.0\textwidth]{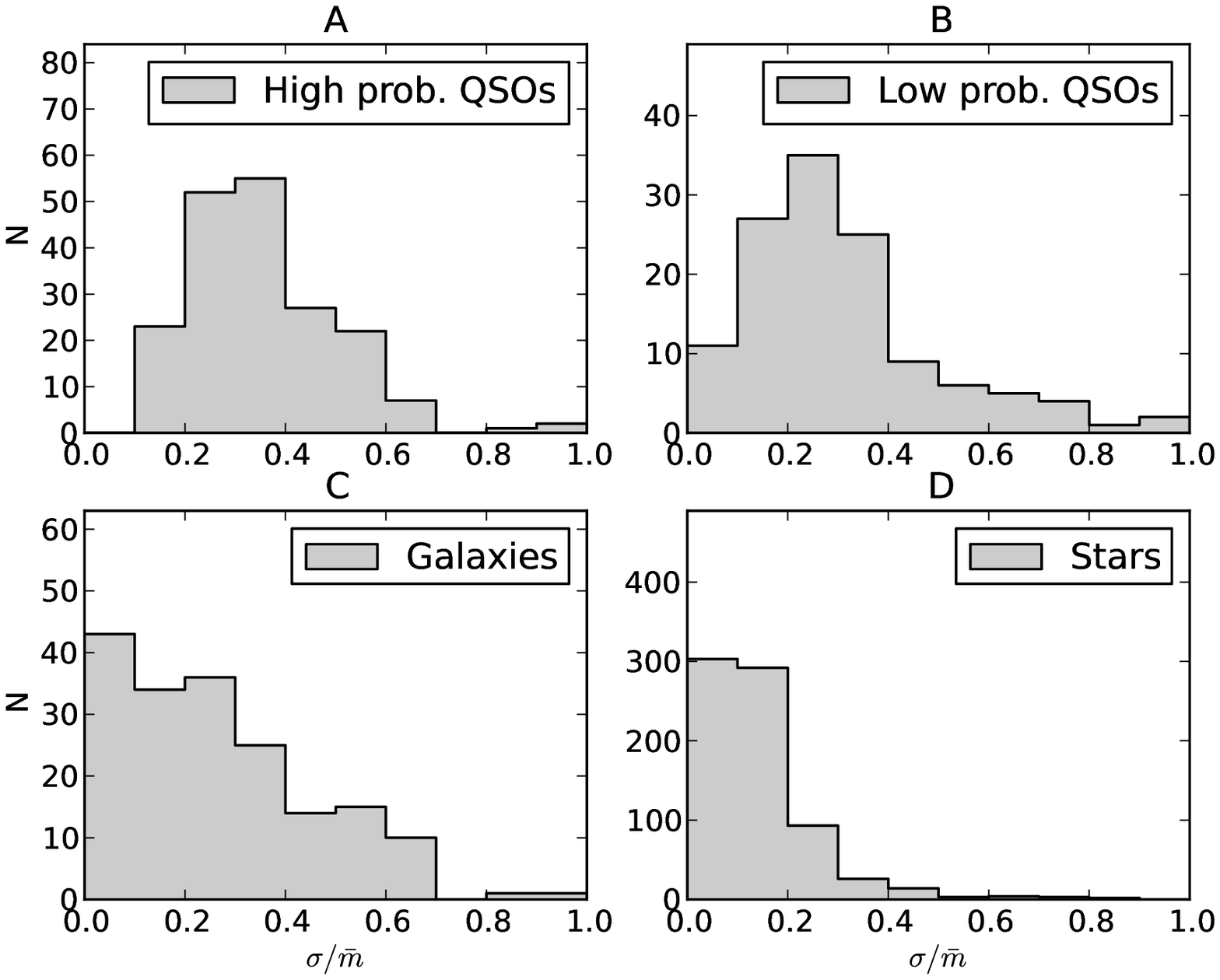} %./MACHO/photoz/plot_type12.py : find '#plot 11 features'
\end{minipage}
\begin{minipage}[c]{8cm}
        \includegraphics[width=1.0\textwidth]{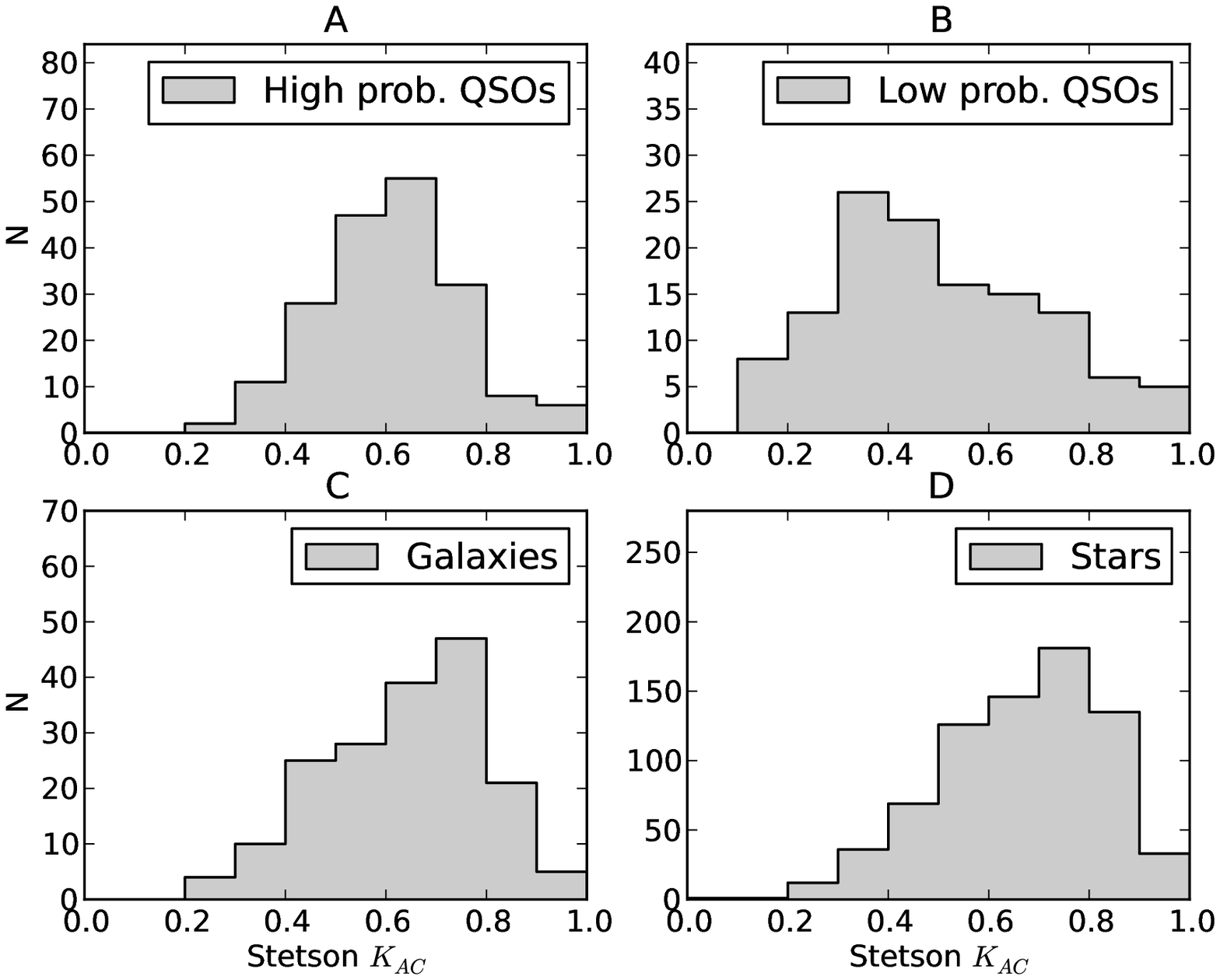}
\end{minipage}
\end{center}
    \caption
           {left side A, B, C and D: Histogram of one of the time series features, $\sigma/\bar{m}$ \citep{Kim2011ApJ}. 
           Galaxies and stars show different distribution from both high and low probability QSOs
           while high and low probability QSOs do not show distinctive differences.
           right side A, B, C and D: Histogram of Stetson $K_{AC}$ \citep{Kim2011ApJ}.
           High probability QSOs show different distribution from low probability QSOs while
           galaxies and stars show almost identical distributions.
           As the histograms show, it seems that the bimodality in the left panel of Figure \ref{fig:photoZ_prob}
           is correlated with the different variability characteristics of each class.
           Further analysis of this bimodality is beyond the scope of this paper. 
	}
    \label{fig:photoZ_feature_hist}
\end{figure*}

The left panel of Figure \ref{fig:photoZ_prob}
also shows similar bimodality as seen in the Figure \ref{fig:SAGE_prob}.
In order to check if there exists 1)  different variability characteristics between QSOs, galaxies and stars, 
and 2)  different variability characteristics between the high and low probability QSO candidates,
we show histograms of two variability features defined in \citet{Kim2011ApJ} in Figure \ref{fig:photoZ_feature_hist}.
The left $2\times2$ sub-panels (left side A, B, C and D)
shows the histogram of $\sigma/\bar{m}$,
where $\sigma$ is the standard deviation and
$\bar{m}$ is  the mean magnitude. In general 
$\sigma/\bar{m}$ is large when a lightcurve has strong variability.
The x-axis is scaled to be between 0 and 1. 
To check if differences exist between high and low probability QSOs (A and B),
we selected two subsets: one  of high ($\ge$80\%) and the other of low ($\le$40\%) probability QSOs.
We included all  galaxies (C) and  stars (D) regardless of their probabilities.
As the left panels show,  galaxies and stars show  different distributions
from the distribution of QSOs that has a peak around $\sim$0.3.
Nevertheless, high and low probability QSOs do not
show different distribution.
The right    $2\times2$  sub-panels (right side A, B, C and D)
 show a different time variability index,
Stetson $K_{AC}$, which is the observation of the distribution of data points 
between the maximum and minimum values of the autocorrelation function 
of a lightcurve \citep{Kim2011ApJ}.
As the panels  show, high probability QSOs (A) show a peak
around 0.6 while low probability QSOs (B) show a peak around 0.4.
Galaxies (C) and stars (D) show peaks around 0.7.
Thus it seems that the bimodality shown in the left panel of Figure \ref{fig:photoZ_prob}
and the different distributions between QSOs, galaxies and stars in Figure \ref{fig:photoZ_prob}
is correlated with the different variability characteristics of the lightcurves.
Further analysis  of this bimodality, requiring careful investigation of many variability 
characteristics and understanding of the selection biases is beyond the scope of this paper.

\begin{figure*}
\begin{center}
\begin{minipage}[c]{8cm}
        \includegraphics[width=1.0\textwidth]{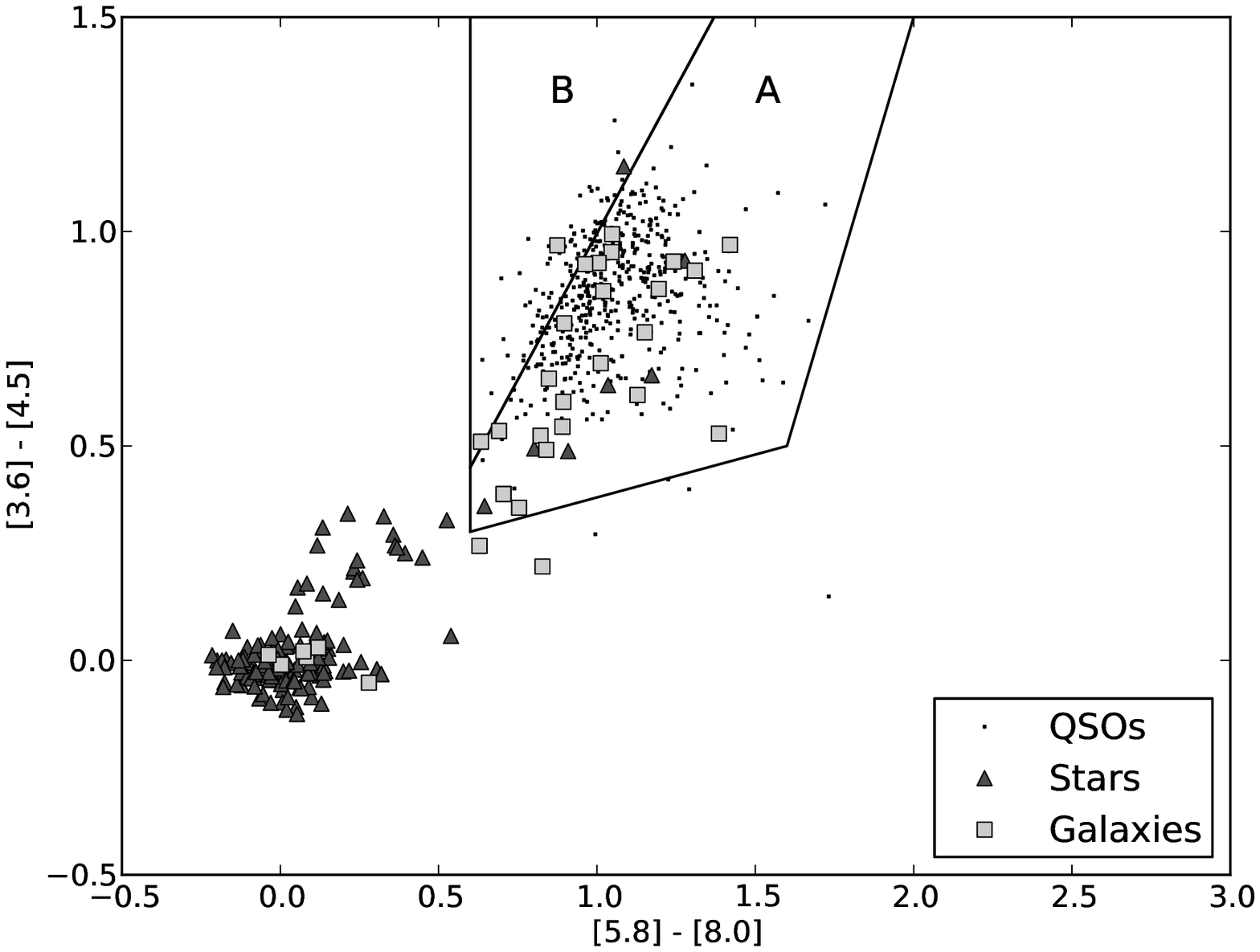} %./MACHO/photoz/plot_photoz_all.py
\end{minipage}
\begin{minipage}[c]{8cm}
        \includegraphics[width=1.0\textwidth]{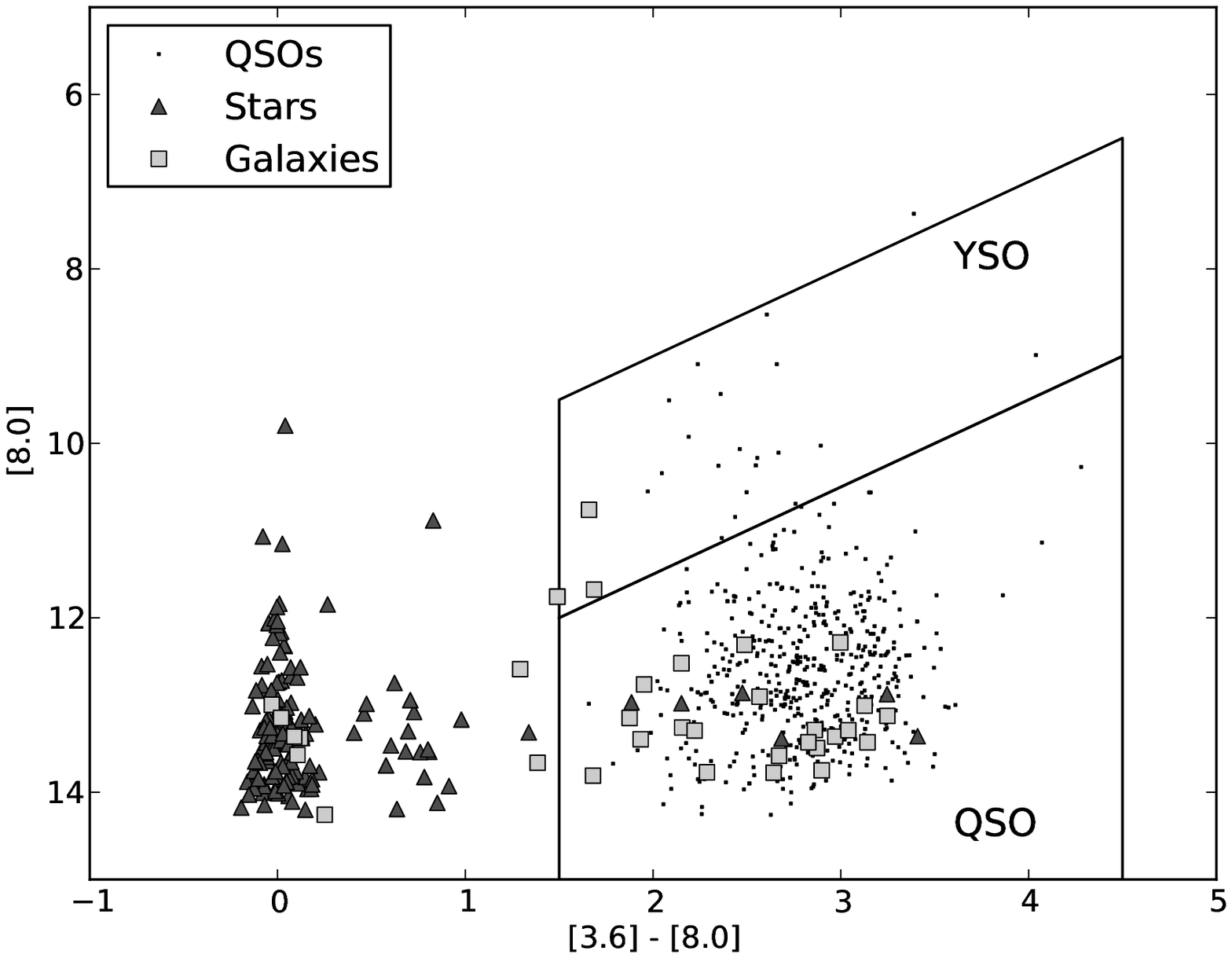}
\end{minipage}
\end{center}
    \caption
           {Mid-IR color-color and color-magnitude diagrams of the QSOs, galaxies and stars
           classified using the photometric redshift code. See Section \ref{sec:photometric_redshift} for details.
           Each axis of the figure is either Spitzer magnitude or color.
           In the left panel, there are 502 QSOs (dots), 33 galaxies (squares) and 145 stars (triangles).
           In the right panel, there are 518 QSOs, 34 galaxies and 145 stars.
           As the figures show, almost all of the QSOs and galaxies are inside the 
           regions (QSO, YSO, A and B), which indicates that all of them are potential QSOs.
           On the other hand, the majority of the stars are outside the regions.
}
    \label{fig:photoZ_sage_can_CCD_CMD}
\end{figure*}

In addition, Figure \ref{fig:photoZ_sage_can_CCD_CMD} shows
the mid-IR colors of QSOs, galaxies and stars.
As the figure shows, almost all of the QSOs (dots)
are inside the four  regions while most of the stars (triangles)
are outside the regions. Galaxies (squares) are either inside or outside the regions.

\subsection{X-ray Luminosity}
\label{sec:Xray_luminosity}

\begin{figure*}
\begin{center}
\begin{minipage}[c]{8cm}
        \includegraphics[width=1.0\textwidth]{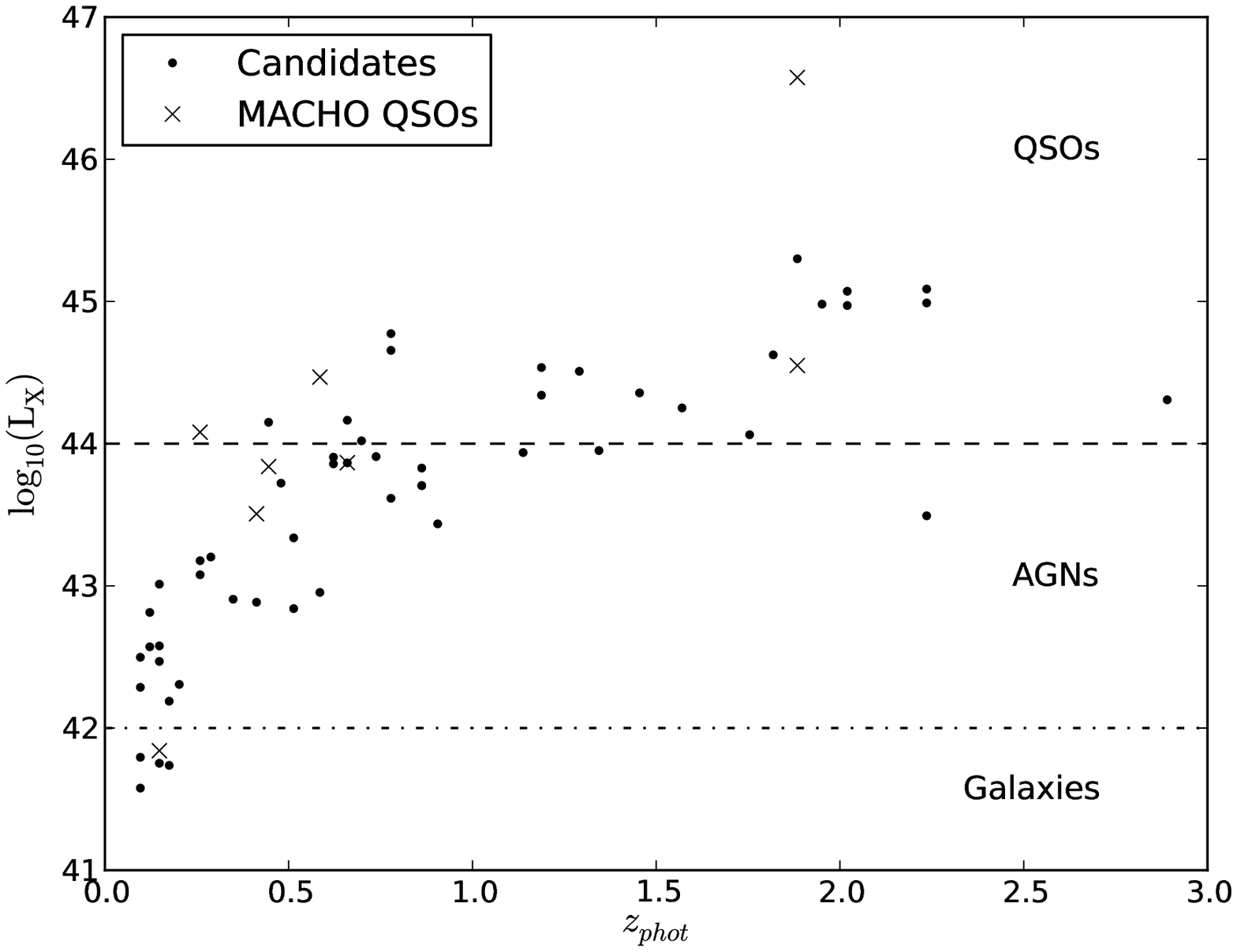} %./MACHO/photoz/plot_lx.py
\end{minipage}
\begin{minipage}[c]{8cm}
        \includegraphics[width=1.0\textwidth]{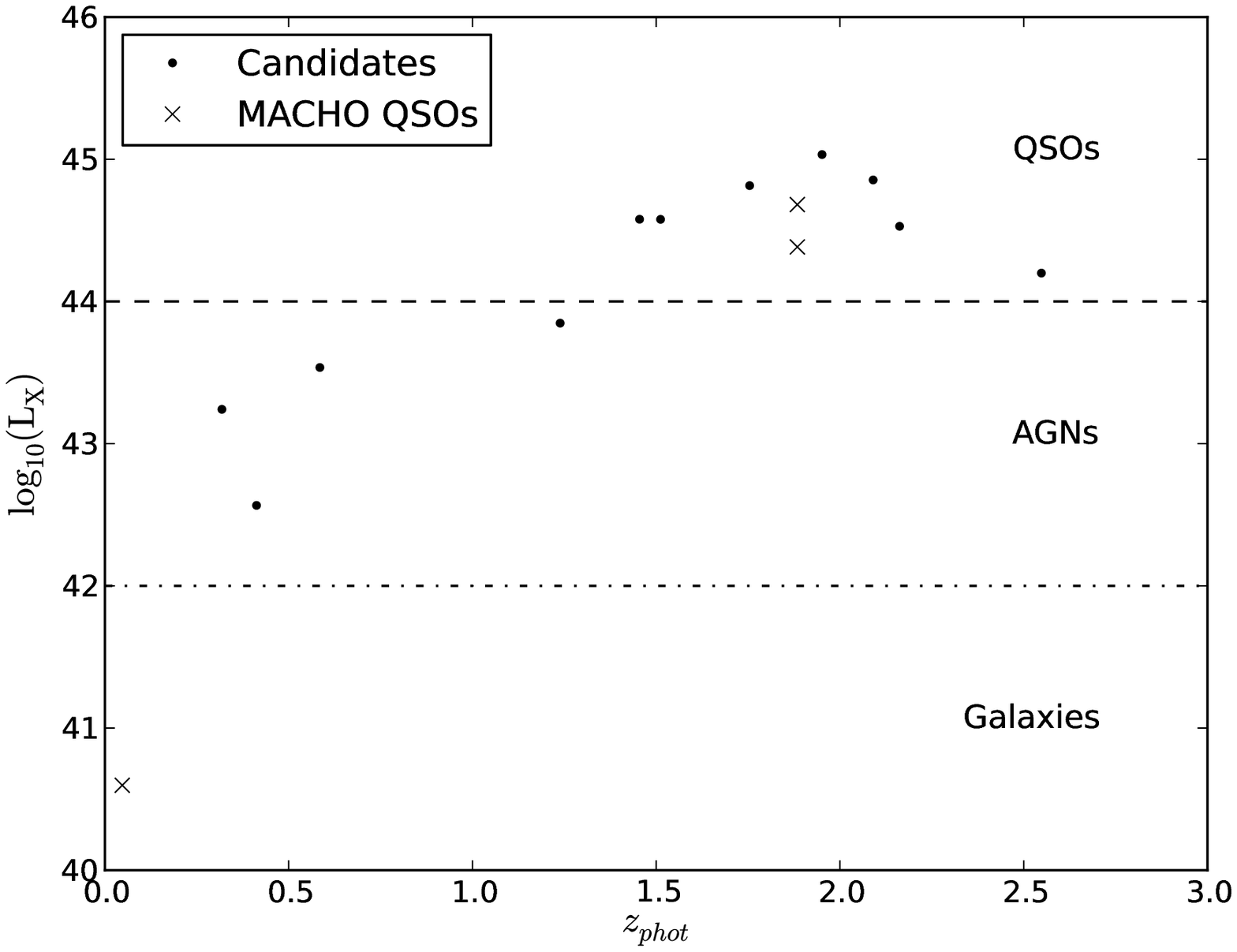}
\end{minipage}
\end{center}
    \caption
           {Scatter plot of the photometric redshifts (x-axis)
	    and the estimated X-ray luminosity, ${\text{log}}{\,\text{L}}_{{\text{X}}}$, (y-axis).
	    The dots are our QSO candidates and the x's are the confirmed MACHO QSOs.
           left: XMM counterparts. 
           right: Chandra counterparts. 
           As the figures show, most of our candidates and MACHO QSOs have 
           ${\text{log}}{\,\text{L}}_{{\text{X}}} > 42$, which indicates they are likely QSOs.
           }
    \label{fig:photoZ_xray}
\end{figure*}

\begin{figure*}
\begin{center}
\begin{minipage}[c]{8cm}
        \includegraphics[width=1.0\textwidth]{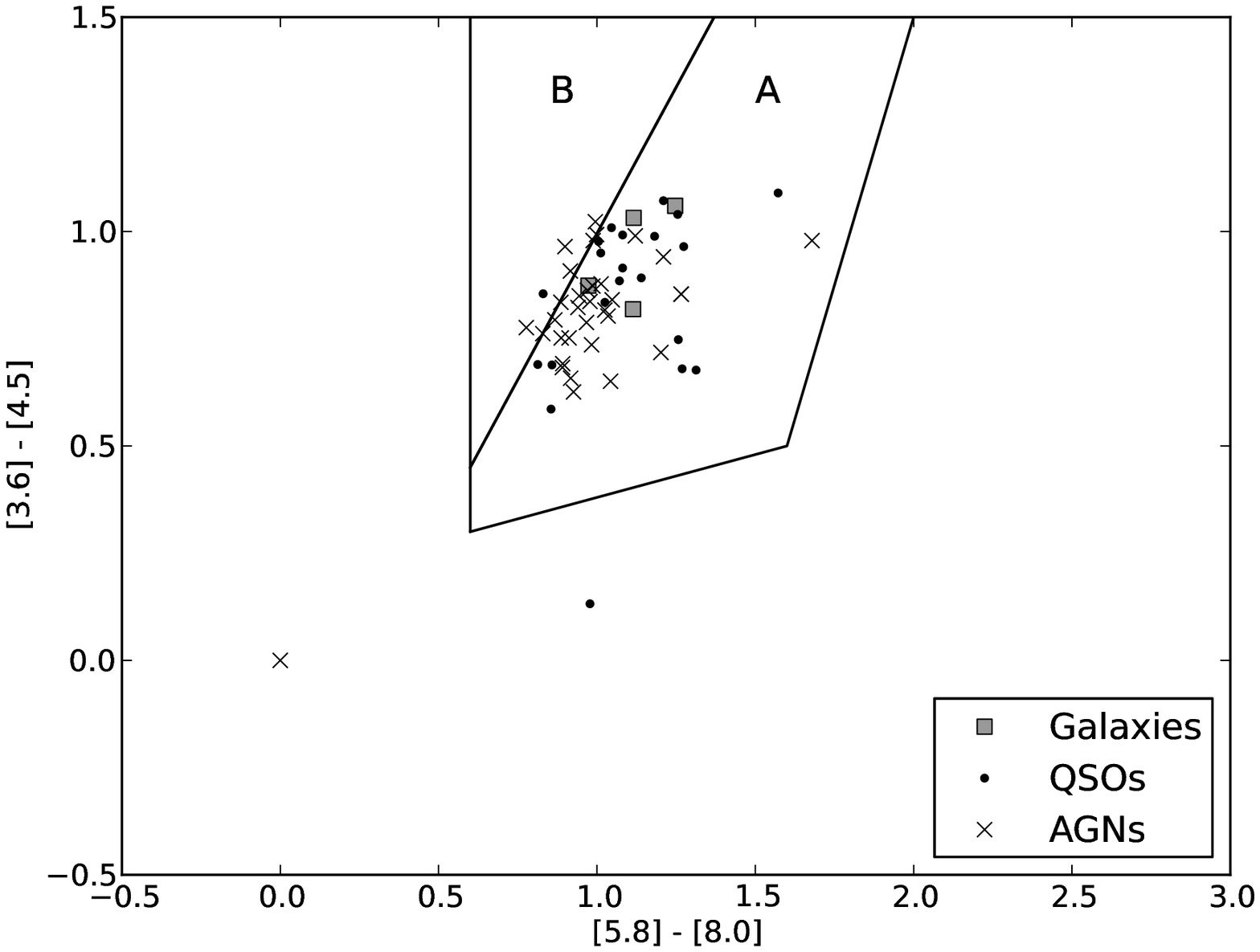} %./MACHO/photoz/plot_lx.py
\end{minipage}
\begin{minipage}[c]{8cm}
        \includegraphics[width=1.0\textwidth]{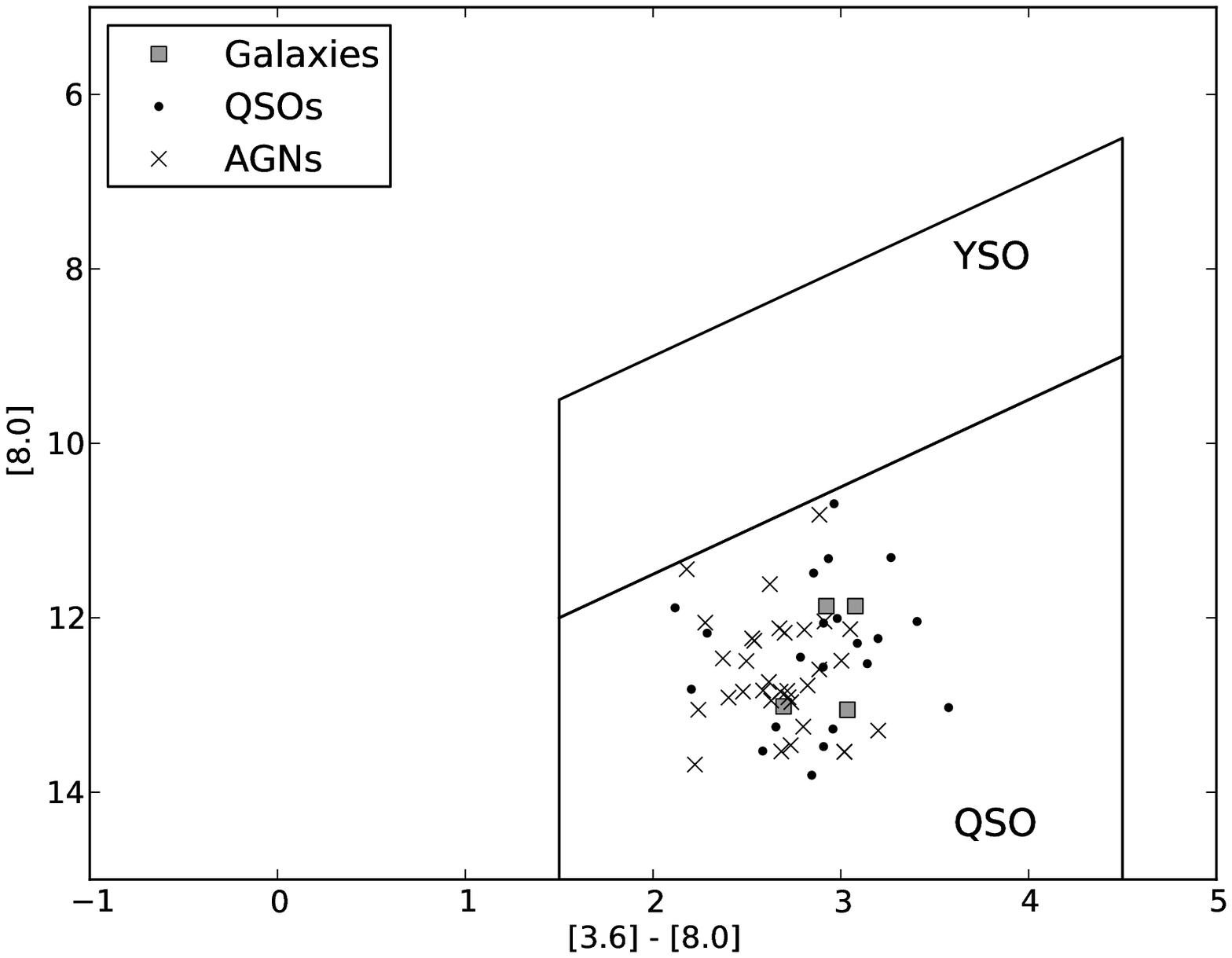}
\end{minipage}
\end{center}
    \caption
           {Mid-IR color-color and color-magnitude diagrams of the QSOs, galaxies and stars
           classified using the X-ray luminosity. See Section \ref{sec:Xray_luminosity} for details.
           Each axis of the figure is either Spitzer magnitude or color.
           As the figures show, almost of the X-ray counterparts are within the QSO and the A region.
           The candidates inside the QSO and the A region are very likely QSOs \citep{Kozlowski2009ApJ}.
}
    \label{fig:photoZ_sage_can_CCD_CMD_xray}
\end{figure*}

In order to estimate the X-ray luminosity,
we  crossmatched the \NCanAll{} QSO candidates 
with two X-ray point source catalogs:
the Chandra X-ray source catalog \citep{Evans2010ApJS} and
the XMM-Newton 2$^{nd}$ Incremental Source catalog \citep{Watson2009AA}.
We searched for the nearest source  within a 5$^{\prime\prime}$ search radius from each candidate.
The majority of the crossmatched counterparts were within
a 3$^{\prime\prime}$ distance from the candidates and 
there were no additional counterparts within a 5$^{\prime\prime}$ 
distance from the candidates.
We found 88 counterparts from either the XMM or Chandra catalogs.

Amongst the 88 counterparts,
64 were fitted with the SED templates mentioned in section \ref{sec:photometric_redshift}
and therefore had estimated photometric redshifts. 
We used the photometric redshifts and X-ray fluxes from the catalogs 
to calculate the X-ray luminosity of each counterpart. 
Figure \ref{fig:photoZ_xray} shows the photometric redshifts (x-axis)
and the estimated X-ray luminosity, ${\text{log}}{\,\text{L}}_{{\text{X}}}$ (y-axis).
In the left panel, we show the 61 XMM counterparts including eight confirmed MACHO QSOs.
The right panel shows 14 Chandra counterparts including three confirmed MACHO QSOs.
Almost all of the candidates (\NCanXray{})
have higher ${\text{log}}{\,\text{L}}_{{\text{X}}}$ than 42.
In addition, six confirmed MACHO QSOs 
and 26 candidates show ${\text{log}}{\,\text{L}}_{{\text{X}}}$ higher than 44.
The candidates showing higher ${\text{log}}{\,\text{L}}_{{\text{X}}}$ than 44 (42)
are likely to be QSOs (AGNs) \citep{Elvis1994ApJS, Persic2004AA}.
The remaining candidates that show lower 
${\text{log}}{\,\text{L}}_{{\text{X}}}$ than 42 are likely to be galaxies.

We show the mid-IR colors of these X-ray counterparts in Figure \ref{fig:photoZ_sage_can_CCD_CMD_xray}.
The classification of QSOs (dots), AGNs (x's) and galaxies (squares)
are based on the X-ray luminosity of the counterparts.

\section{High Confidence QSO Candidate Selection Using Support Vector Machines}
\label{sec:finalCatalog}

\subsection{Support Vector Machine}

SVM  (Support Vector Machine, \citealt{Boser1992}) is a supervised machine learning algorithm
that trains a two-class classification model using samples of two known classes (i.e. training set).
SVM is currently one of the best classification methods in machine learning. 
The classifier of a SVM defines a linear hyperplane that separates two
classes in a training data.
To select a unique hyperplane among the set of
possible hyperplanes that separate the data, SVM chooses the hyperplane which
maximizes the margin between the two classes, 
and is therefore often called
the {\em{maximum margin separator}}.
SVM is also able to separate non-linearly separable  classes
by using a kernel function (e.g. a polynomial kernel or a radial basis kernel)
transforming non-linear feature spaces into linear feature spaces.
The hypothesis of SVM has the form:

\begin{equation}
\label{eq:svmhyp}
Class(z) = \mbox{sign}(\sum_i \alpha_i y_i  K(z,x_i) - b)
\end{equation}

{\noindent}where 
$i$ are the indices for training set examples,
$x_i$ are the examples,
$y_i$ are the labels,
$z$ is the example that we are predicting the label for, 
$K(z,x_i)$ is a kernel function,
and $b$ is a threshold. The $\alpha_i$ are the parameters learned
by the training procedure.
Despite the mapping to a potentially
high dimensional space using a kernel function, 
the maximum margin criterion leads to automatic
capacity control and thus avoids overfitting.

Compared to neural networks, SVMs provide a flexible classification
model, avoids the problems of local minima, and reduces the need for parameter
tuning. For an overview, discussion and practical details,
see \citet{Cristianinic2000, BennettC00, libsvmguide, Kim2011ApJ} and references therein.
Because standard SVM can only solve a two-class problem,
\citet{Scholkopf2001} proposed a method to solve 
one-class classification problems using SVM.
In brief, they define the origin as the second class
and separate the one class from the origin using SVM.
For details about the method, see \citet{Scholkopf2001, Manevitz2002}.

\subsection{Training a one-class SVM to Select High Confidence QSO Candidates}

\begin{figure*}
\begin{center}
        \includegraphics[width=0.75\textwidth]{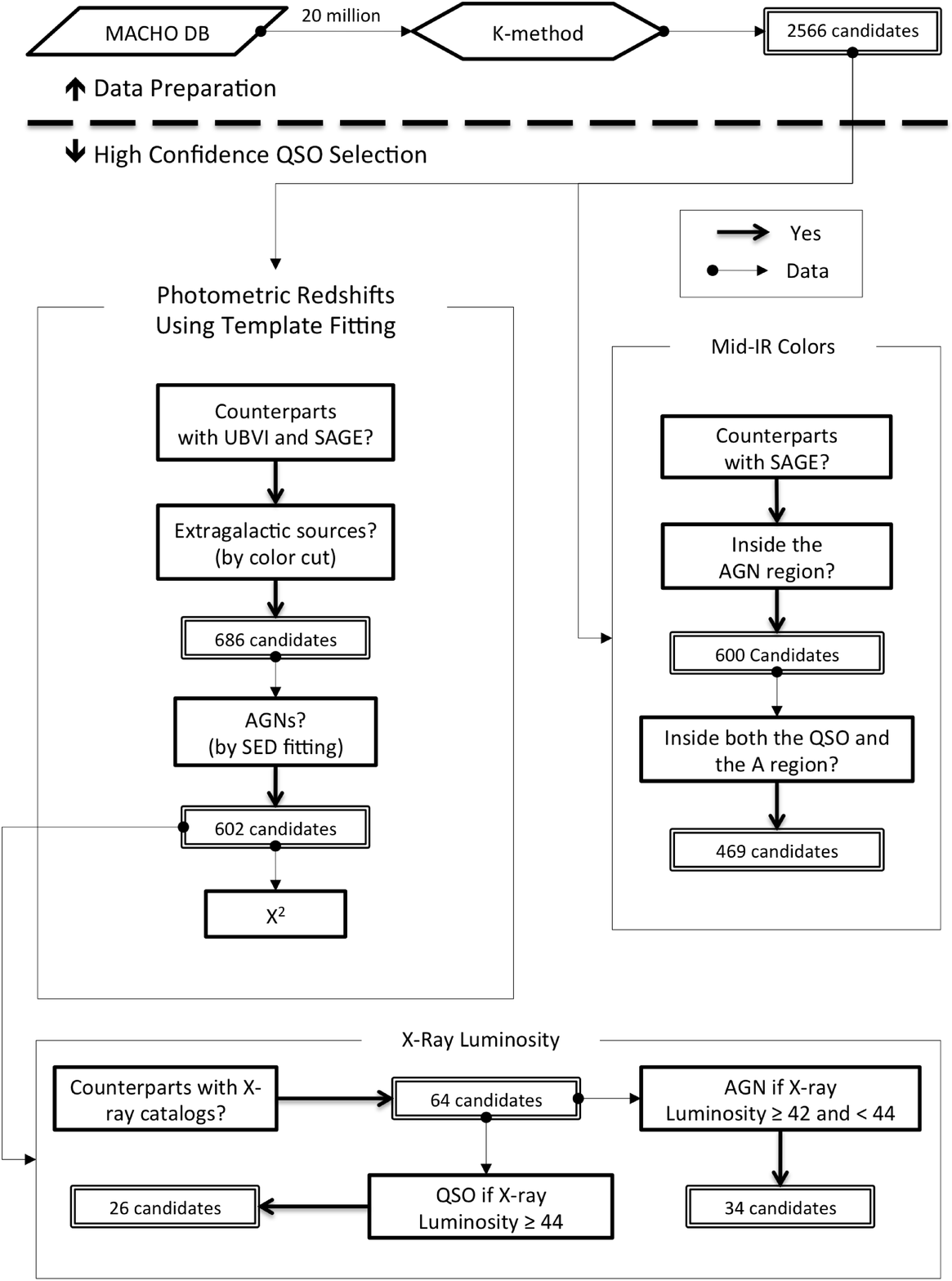} %.Tree_v2.pptx in the dropbox folder. save it to png (:use big canvas) and convert it to eps using 'convert'.
    \caption
           {Illustration of the processes that we used to select hc-QSOs. 
           The rectangles with bold borderlines are the diagnostics. 
           At most of the diagnostics, we determined 
           if the candidates are likely to be QSOs (solid line arrows).
           The thin arrows show the data flow. The double-lined rectangles
           show the number of candidates.}
    \label{fig:Tree}
\end{center}
\end{figure*}

We employed the one-class SVM classification method
to select high confidence QSO candidates
because we do not have negative examples (i.e. non-QSO training set).
We used a linear kernel rather than a polynomial kernel
or a radial basis kernel because we empirically
found that using other kernels did not improve classification results.
To train a model, we first defined the diagnostics results as feature vectors.
Table \ref{tab:feature_vectors} summarizes the feature vectors.
When we could not determine a feature value 
due to the nonexistence of counterpart with either the Spitzer SAGE,
UBVI and X-ray catalogs, we assigned zero to the corresponding feature.
Figure  \ref{fig:Tree} outlines the calculation of the
diagnostics and the number of candidates for which 
the diagnostics are available.
As mentioned above, we started with the \NCanAll{} QSO candidates selected
using the \QSOKim{}
(`Data Preparation' panel in the figure).
The diagnostics applied to these candidates
are shown in the `High Confidence QSO Selection' panel.
We also show the number of QSO candidates
after the diagnostics (double-lined rectangles).

\begin{figure}
\begin{center}
        \includegraphics[width=0.45\textwidth]{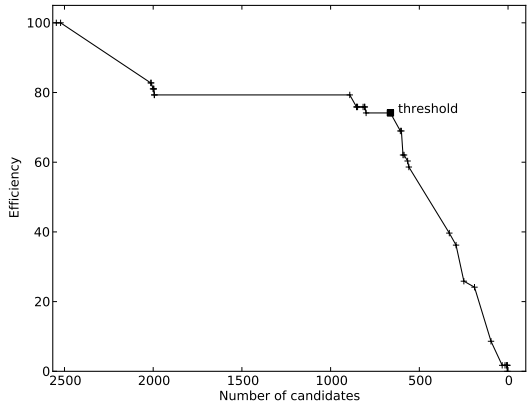} %./MACHO/ML_oneclass/threshold.py
    \caption
           {Efficiency versus number of selected QSO candidates as a function of 
           the SVM threshold, $b$. The black square shows the final threshold we adopted.}
    \label{fig:threshold}
\end{center}
\end{figure}

We trained a one-class SVM model
using these features.\footnote{We 
used the \href{http://www.csie.ntu.edu.tw/~cjlin/libsvm}{\fontfamily{pcr}\selectfont LIBSVM package} \citep{Chang2001}.}  
We then tuned the model by adjusting the threshold, $b$, in order to:
1) obtain the highest efficiency based on the confirmed 58 MACHO QSOs,
and 2) minimize the number of selected QSO candidates, 
which reduces the number of false positives as well.
Figure \ref{fig:threshold} shows the efficiency and
the number of candidates as a function of $b$.
The black square shows the threshold we finally adopted.
Using the determined threshold, the trained model showed 74\% efficiency.
We applied the tuned model to the \NCanAll{} QSO candidates
and selected \NCanStrong{} QSO candidates (i.e. hc-QSOs).

Table \ref{tab:can_table} shows a few important parameters for some of the QSO candidates.
The entire parameters of the \NCanAll{} QSO candidates are published in the electronic edition of this manuscript.
We also provide catalogs and lightcurves of all the candidates
at \href{http://timemachine.iic.harvard.edu/coati/QSOs/index.html}
{http://timemachine.iic.harvard.edu/coati/QSOs}.

\begin{table*}
\begin{center}
\caption{Feature Vectors \label{tab:feature_vectors}}
\begin{tabular}{cccccc}
\tableline\tableline
mid-IR & extragalactic sources/stars & SED fitting & $\chi^{2}$ & Chandra & XMM \\
\tableline
\tableline

no CP\footnote{no counterpart.} : 0 & no CP : 0 & no CP : 0  & no CP : 0 & no CP : 0 & no CP : 0 \\
 inside any of the four regions : 1 & stars : 1 & galaxies : 1 & $\chi^{2}$ value\footnote{$\chi^{2}$ is from the SED fitting.} & galaxies : 1 & galaxies : 1 \\
inside both the QSO and A region : 2 & extragalactic sources : 2 & AGNs : 2 & & AGNs : 2 & AGNs : 2 \\
& & & & QSOs : 3 & QSOs : 3 \\

\tableline

\end{tabular}
\end{center}
\end{table*}

\begin{table}
\begin{center}
\caption{Several Important Parameters of the QSO candidates \label{tab:can_table}}
\begin{tabular}{rcccc}
\tableline\tableline
MACHO ID & RA & Dec & V & hc-QSO\footnote{1: high confidence QSO candidate} \\
                 & (in degree)     &  (in degree)     &  (mag)   &  \\
\tableline
\tableline

 11.8747.1083  & 83.52708   & -70.62689  & 18.98   & 1     \\
 11.8753.346   & 83.66207   & -70.20544  & 18.72   &       \\
 11.8984.29    & 83.89623   & -70.89459  & 18.16    &       \\
 11.8989.258   & 84.04636   & -70.61672  & 18.67   &       \\
 11.8994.1323  & 83.91927   & -70.25463  & 19.27   & 1     \\
 11.9349.1074  & 84.53299   & -70.81329  & 20.26   & 1     \\
 11.9353.1217  & 84.52798   & -70.50399  & 19.52   &       \\
 12.10679.528  & 86.51372   & -70.85550   & 18.99   &       \\
 13.5834.232   & 79.19451   & -71.16704  & 19.57    &       \\
 13.6446.758   & 80.00723   & -70.74329  & 20.32   &       \\
 13.6448.3756  & 80.03121   & -70.59326  & 19.04   &       \\
 13.6560.555   & 80.25070    & -71.21474  & 19.68    &       \\

\tableline

\end{tabular}
\end{center}
Note: This table is published in its entirety in the electronic edition of this manuscript. A portion is shown here for guidance regarding its form and content.
%the entire electronic table is generated using http://aas.org/journals/authors/common_instruct#_Toc6.1 and http://dopey.mcmaster.ca/MRT/upload.html

\end{table}

\section{Crossmatching with Newly Discovered QSOs by Koz{\L}owski (2011)}
\label{sec:koz_QSOs}

Recently, \citet{Kozlowski2011arXiv} selected QSO candidates
using  mid-IR colors, X-ray emission 
and/or optical variability in the OGLE lightcurve database \citep{Udalski2008AcA}.
For the  variability selection, they used
the DRW (a Damped Random Walk) model of lightcurves \citep{Kelly2009ApJ, Kozlowski2010ApJ}
and then applied several cuts including
magnitude, model fitting accuracy\footnote{The likelihood ratio between
the best fitting model and a white noise model.}, 
slope of a structure function, amplitude and time scale of lightcurve variations.
They then visually examined all the lightcurves of the candidates and removed 
about 96\% of lightcurves ($\sim$23,000) from the final list. Most of false positives
were the `ghost' variable objects caused by photometric defects.
They finally observed 845 QSO candidates using 
AAT/AAOmega\footnote{AAT: Anglo-Australian Telescope, AAOmega: the AAT multi-purpose fiber-fed spectrograph 
\citep{Sharp2006SPIE}.} and confirmed 169 QSOs 
including 25 previously known QSOs\footnote{
18 of them are on the confirmed MACHO QSO list and seven of them
are not on the confirmed MACHO QSO list.} (i.e. 144 newly discovered QSOs)
in the four $\sim$3 deg$^{2}$ field near the LMC center.
They also provided the list of remaining 676 objects.
Among these 676 objects,  they  confirmed that 275 are  non-QSOs,
including young stellar objects (YSOs), red stars, blue stars, Be stars and 
planetary nebulae.\footnote{The remaining  sources
had undetermined classification.}

To estimate the efficiency and the false positive rate of our selection method,
we first crossmatched the 151 discovered QSOs\footnote{144
newly discovered QSOs and seven previously known QSOs
that are not on the confirmed MACHO QSO list.}
and 275 confirmed non-QSOs (i.e. false positives) with the entire MACHO LMC lightcurve database.
We searched the nearest MACHO LMC source
within a 3$^{\prime\prime}$ search radius.
Out of 151 QSOs and 275 non-QSOs, 64 and 122 were crossmatched with the MACHO sources.
Note that, only 46 out of 64 were selected using variability characteristics
in the OGLE-III lightcurves \citep{Kozlowski2011arXiv}.

Among these 46 QSOs, 20 are in the hc-QSO list (hereinafter, c-QSOs)
and 26 are not in the hc-QSO list (hereinafter, cn-QSOs),
which gives us 43\%  efficiency. 
It is worth mentioning that the yield of QSO candidates from \citet{Kozlowski2011arXiv}
selected using only variability based on the DRW model was 
7\%.

\begin{figure}
\begin{center}
       \includegraphics[width=0.45\textwidth]{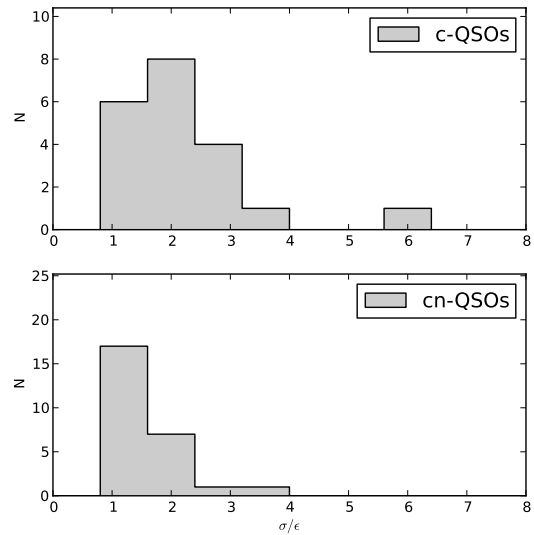} % ./MACHO/LMC_koz/plot_LCs.py
\end{center}
    \caption{Histogram of
the ratio between the photometric uncertainty and amplitude, $\sigma / \epsilon$, of 
c-QSOs (the top panel) and cn-QSOs (the bottom panel). See the text for details about c-QSOs and cn-QSOs.
Small $\sigma / \epsilon$ means that the photometric uncertainty is too large to detect variability.
c-QSOs show relatively larger $\sigma / \epsilon$ than cn-QSOs, which means that c-QSOs are more detectable 
than cn-QSOs using variability.}
    \label{fig:std_errors}
\end{figure}

\begin{figure}
\begin{center}
       \includegraphics[width=0.45\textwidth]{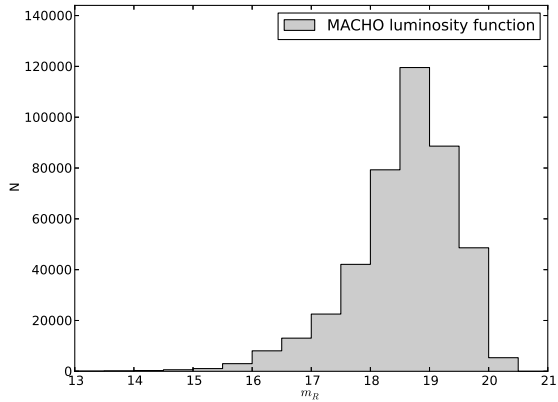} % ./MACHO/LMC_koz/plot_luminosity_function_macho.py
\end{center}
    \caption{Luminosity function of MACHO R magnitude from one MACHO field.
    The x-axis is MACHO R magnitude and the y-axis is the number of MACHO sources.
    As the figure shows, the limiting R magnitude is around 19 $\sim$ 19.5.}
    \label{fig:MACHO_luminosity_function}
\end{figure}

Despite of the fact that these 46 QSOs were determined to be variable objects
based on the optical OGLE-III lightcuves,
some of them do not show strong variability in the MACHO lightcurves
because of 1) the difference of the limiting magnitudes of the two survey,
and 2) the photometric uncertainty of the MACHO lightcurves.
For instance, we found that 11 of cn-QSOs
are fainter than 19 MACHO R magnitude ($m_{R}$)
while only two of c-QSOs are fainter than 19 $m_{R}$,
which is around a limiting magnitude of MACHO survey (Figure \ref{fig:MACHO_luminosity_function}).
Thus it is  likely that the \QSOKim{} using variability
was not able to detect some of the QSOs
due to the large photometric uncertainty and thus weak variability.
Figure \ref{fig:std_errors} shows the histogram of
the ratio between the average photometric uncertainty 
and standard deviation (i.e. amplitude), $\sigma / \epsilon$, 
of the lightcurves of c-QSOs and cn-QSOs.
Small $\sigma / \epsilon$ means that 
the photometric uncertainty
is relatively larger than the amplitude of the lightcurve,
which implies that it is rather hard to detect its variability.
As the figure shows, c-QSOs have relatively larger
$\sigma / \epsilon$ than cn-QSOs, which means c-QSOs are more detectable 
than cn-QSOs using their variability. 
$\sigma$ is one of the time variability features that the \QSOKim{} used.

\begin{figure}
\begin{center}
       \includegraphics[width=0.45\textwidth]{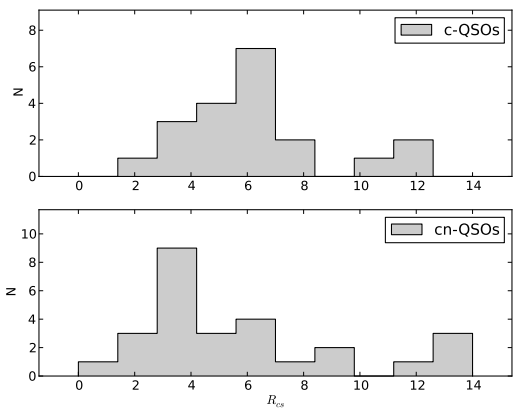} % ./MACHO/LMC_koz/plot_features_LCs.py
\end{center}
    \caption{Histogram of $R_{cs}$ of c-QSOs (the top label) and cn-QSOs (the bottom panel).
    c-QSOs and cn-QSOs show different distribution. See the text for details.}
    \label{fig:cusum_koz}
\end{figure}

In Figure \ref{fig:cusum_koz}, we show an alternative way of seeing variability characteristic
of a lightcurve by borrowing one example of 
the time series features, $R_{cs}$ \citep{Ellaway1978}, used in the \QSOKim{}.
$R_{cs}$, the range of a cumulative sum, is typically large for the variables showing non-periodic and strong variability, 
and is small for periodic variables or non-variables.
As the figure shows, the histogram of c-QSOs (the top panel) has a peak around 6
while the histograms of cn-QSOs shows a peak around 3 (the bottom panel).

\begin{figure*}
\begin{center}
\begin{minipage}[c]{15cm}
       \includegraphics[width=1\textwidth]{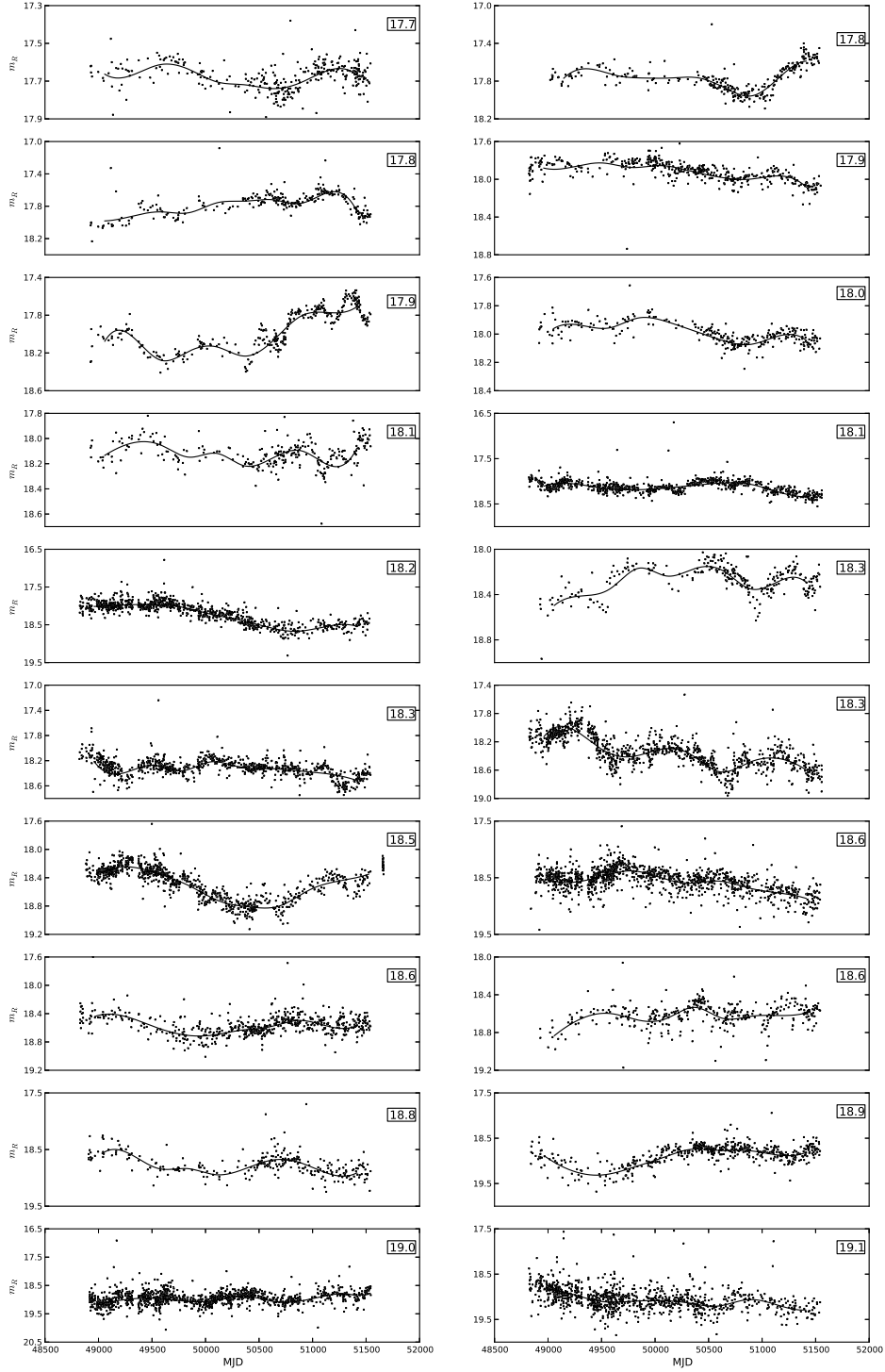} % ./MACHO/LMC_koz/plot_LCs_2.py
\end{minipage}
\end{center}
    \caption
    	{MACHO B band lightcurves of c-QSOs.
	The x-axis is MJD and the y-axis is MACHO R magnitude ($m_{R}$).
	The solid lines are the smoothed spline lightcurves.
	The small boxes inside each panel show the average $m_{R}$.
	As the figure shows, almost all the lightcurves show strong variability
	regardless of their magnitudes.}
    \label{fig:LCs_in_our_can}
\end{figure*}

\begin{figure*}
\begin{center}
\begin{minipage}[c]{15cm}
       \includegraphics[width=1\textwidth]{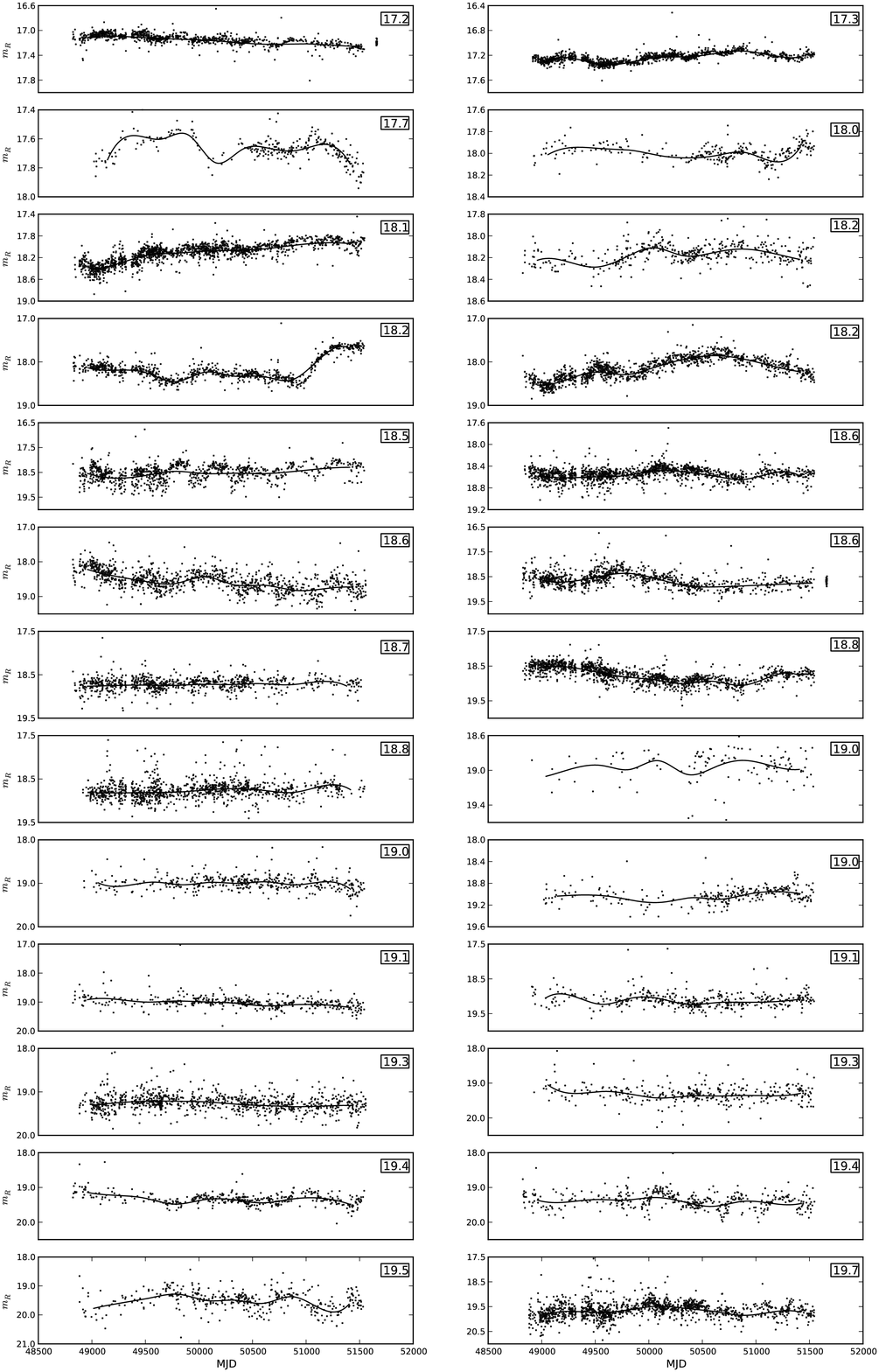} % ./MACHO/LMC_koz/plot_LCs_2.py
\end{minipage}
\end{center}
    \caption{MACHO B band lightcurves of cn-QSOs.
	When compared to the lightcurves shown in Figure \ref{fig:LCs_in_our_can},
	these lightcurves show relatively weaker variability. Moreover, there are
	a lot more fainter lightcurves than the lightcurves in Figure \ref{fig:LCs_in_our_can}.
	}
    \label{fig:LCs_not_in_our_can}
\end{figure*}

In addition, we show the MACHO lightcurves of the 
20 c-QSOs and 26 cn-QSOs in 
Figure \ref{fig:LCs_in_our_can} and Figure \ref{fig:LCs_not_in_our_can}.
As Figure \ref{fig:LCs_in_our_can} shows, most of the c-QSOs show strong variability.
On the other hand, Figure \ref{fig:LCs_not_in_our_can} shows
that most of the cn-QSOs fainter than 19 $m_{R}$ show relatively weaker
variability than the variability of c-QSOs. 
Only cn-QSOs brighter than 19 $m_{R}$ shows strong variability
comparable to that of c-QSOs.

According to Figure \ref{fig:std_errors}, \ref{fig:cusum_koz}, \ref{fig:LCs_in_our_can} and
\ref{fig:LCs_not_in_our_can}, it seems that the main reason for the non-detection of
QSOs is the relatively weaker variability.
Thus if we ignore some of the QSOs 
showing weak variability, our efficiency would be higher than 43\%.
For instance, if we ignore the 11 cn-QSOs
fainter than 19 $m_{R}$, our efficiency increases to 57\%.

In the case of the false positives, only two out of 
122 confirmed non-QSOs are inside the hc-QSO list,
which gives 0.3\% false positive rate.
The two false positives are YSOs.
We examined the MACHO lightcurves of them
and confirmed that they show strong variability.
Note that \citet{Kozlowski2011arXiv}  monitored 12 deg$^{2}$ fields around the LMC
that are mostly inside the 40 deg$^{2}$ MACHO LMC fields.
Given that our QSO candidates are uniformly distributed around the LMC,
we would have about one third number of the hc-QSOs ($12 / 40 $) 
inside the fields that \citet{Kozlowski2011arXiv} monitored.
In such case, the false positive rate is about 1\%.
However the true false positive rate would be higher than 1\%
because  \citet{Kozlowski2011arXiv} did not monitor all the sources in the fields,
which means some of our QSO candidates are not in their list. 
Nevertheless, these 122 non-QSOs were selected not only by variability 
but also by mid-IR colors and X-ray emission.
Thus it seems that  our method is successful to 
exclude any type of false positives, which is crucial for
the selection of QSO candidates from massive astronomical databases 
such as Pan-STARRS \citep{Kaiser2004SPIE} and LSST \citep{Ivezic2008arXiv}
due to: 1) the enormous amount of data, which thus could yield huge number of false positives,
and 2) the high cost of spectroscopic observations for such deep and wide field surveys.

\section{Summary}
\label{sec:summary}

In this paper, we presented \NCanStrong{} high confidence QSO candidates, in the LMC fields.
We first selected \NCanAll{} QSO candidates based on the time variability of 
MACHO B and R band lightcurves in the MACHO LMC ligtcurve database 
using the method of \citet{Kim2011ApJ}.
We then applied multiple diagnostics such 
as mid-IR color, photometric redshift and X-ray luminosity to these QSO candidates.
Using the diagnostics outputs,
we trained a one-class SVM model to discriminate
high confidence QSO candidates.
We finally applied the trained model to
the original candidates and selected \NCanStrong{} QSO candidates.

To estimate the yield and false positive rate of the final  list,
we crossmatched them with recently confirmed QSOs 
and non-QSOs in the LMC field \citep{Kozlowski2011arXiv}.
As a result, we found that the yield is higher than $43\%$. 
It is worth mentioning that the yield of the 
QSO candidates selected using the `damped random work' model \citep{Kelly2009ApJ}
is 7\% \citep{Kozlowski2011arXiv} .
In the case of the false positive rate,
we found that there are only a few confirmed non-QSOs in our  list,
which is less than 1\% false positive rate.
Thus this set could be used as a target set potential
for spectroscopic survey to maximize the yield.
This is important because the spectroscopic observations for relatively faint objects 
such as the QSO candidates in dense- and wide-field area 
around the LMC is extremely expensive.
We are planning to use the confirmed QSOs and confirmed non-QSOs 
to improve our QSO selection method. This work will be separately published in the near future.

We will apply our method to the MACHO SMC/bulge database
and the Pan-STARRS MDF (Medium Deep Field) time series database
to further select QSO candidates and thus increase the collection
of QSO lightcurves.

\section*{Acknowledgements}

We thank L. Mylonadis for helpful comments.
The analysis in this paper has been done using the \href{http://hptc.fas.harvard.edu/}{Odyssey cluster} 
supported by the FAS Research Computing Group at \href{http://harvard.edu/}{Harvard}.
This work has been supported by NSF grant IIS-0803409.
This research has made use of the \href{http://simbad.u-strasbg.fr/simbad/}{SIMBAD} 
database, operated at CDS, Strasbourg, France.

\vspace{0.1cm}
\bibliography{KimQSO2011}{}

\begin{thebibliography}{46}
\expandafter\ifx\csname natexlab\endcsname\relax\def\natexlab#1{#1}\fi

\bibitem[{Bennett \& Campbell(2000)}]{BennettC00}
Bennett, K.~P., \& Campbell, C. 2000, SIGKDD Explorations, 2, 1

\bibitem[{{Boser} {et~al.}(1992){Boser}, {Guyon}, \& {Vapnik}}]{Boser1992}
{Boser}, B.~E., {Guyon}, I.~M., \& {Vapnik}, V.~N. 1992, in Proceedings of the
  fifth annual workshop on Computational learning theory, COLT '92 (New York,
  NY, USA: ACM), 144--152

\bibitem[{{Bower} {et~al.}(2006){Bower}, {Benson}, {Malbon}, {Helly}, {Frenk},
  {Baugh}, {Cole}, \& {Lacey}}]{Bower2006MNRAS}
{Bower}, R.~G., {Benson}, A.~J., {Malbon}, R., {Helly}, J.~C., {Frenk}, C.~S.,
  {Baugh}, C.~M., {Cole}, S., \& {Lacey}, C.~G. 2006, MNRAS, 370, 645

\bibitem[{{Chang} \& {Lin}(2001)}]{Chang2001}
{Chang}, C.~C., \& {Lin}, C.~J. 2001, LIBSVM : a library for support vector
  machines, http://www.csie.ntu.edu.tw/$\sim$cjlin/libsvm

\bibitem[{{Cristianini} \& {Shawe-Taylor}(2000)}]{Cristianinic2000}
{Cristianini}, N., \& {Shawe-Taylor}, J. 2000, An Introduction to Support
  Vector Machines (Cambridge: Cambridge Univ. Press)

\bibitem[{{Dobrzycki} {et~al.}(2005){Dobrzycki}, {Eyer}, {Stanek}, \&
  {Macri}}]{Dobrzycki2005AA}
{Dobrzycki}, A., {Eyer}, L., {Stanek}, K.~Z., \& {Macri}, L.~M. 2005, A\&A,
  442, 495

\bibitem[{{Eisenhardt} {et~al.}(2004){Eisenhardt}, {Stern}, {Brodwin}, {Fazio},
  {Rieke}, {Rieke}, {Werner}, {Wright}, {Allen}, {Arendt}, {Ashby}, {Barmby},
  {Forrest}, {Hora}, {Huang}, {Huchra}, {Pahre}, {Pipher}, {Reach}, {Smith},
  {Stauffer}, {Wang}, {Willner}, {Brown}, {Dey}, {Jannuzi}, \&
  {Tiede}}]{Eisenhardt2004ApJS}
{Eisenhardt}, P.~R., {Stern}, D., {Brodwin}, M., {Fazio}, G.~G., {Rieke},
  G.~H., {Rieke}, M.~J., {Werner}, M.~W., {Wright}, E.~L., {Allen}, L.~E.,
  {Arendt}, R.~G., {Ashby}, M.~L.~N., {Barmby}, P., {Forrest}, W.~J., {Hora},
  J.~L., {Huang}, J., {Huchra}, J., {Pahre}, M.~A., {Pipher}, J.~L., {Reach},
  W.~T., {Smith}, H.~A., {Stauffer}, J.~R., {Wang}, Z., {Willner}, S.~P.,
  {Brown}, M.~J.~I., {Dey}, A., {Jannuzi}, B.~T., \& {Tiede}, G.~P. 2004, ApJS,
  154, 48

\bibitem[{{Ellaway}(1978)}]{Ellaway1978}
{Ellaway}, P. 1978, Electroencephalography and Clinical Neurophysiology, 45,
  302

\bibitem[{{Elvis} {et~al.}(1994){Elvis}, {Wilkes}, {McDowell}, {Green},
  {Bechtold}, {Willner}, {Oey}, {Polomski}, \& {Cutri}}]{Elvis1994ApJS}
{Elvis}, M., {Wilkes}, B.~J., {McDowell}, J.~C., {Green}, R.~F., {Bechtold},
  J., {Willner}, S.~P., {Oey}, M.~S., {Polomski}, E., \& {Cutri}, R. 1994,
  ApJS, 95, 1

\bibitem[{{Evans} {et~al.}(2010){Evans}, {Primini}, {Glotfelty}, {Anderson},
  {Bonaventura}, {Chen}, {Davis}, {Doe}, {Evans}, {Fabbiano}, {Galle}, {Gibbs},
  {Grier}, {Hain}, {Hall}, {Harbo}, {(Helen He}, {Houck}, {Karovska},
  {Kashyap}, {Lauer}, {McCollough}, {McDowell}, {Miller}, {Mitschang},
  {Morgan}, {Mossman}, {Nichols}, {Nowak}, {Plummer}, {Refsdal}, {Rots},
  {Siemiginowska}, {Sundheim}, {Tibbetts}, {Van Stone}, {Winkelman}, \&
  {Zografou}}]{Evans2010ApJS}
{Evans}, I.~N., {Primini}, F.~A., {Glotfelty}, K.~J., {Anderson}, C.~S.,
  {Bonaventura}, N.~R., {Chen}, J.~C., {Davis}, J.~E., {Doe}, S.~M., {Evans},
  J.~D., {Fabbiano}, G., {Galle}, E.~C., {Gibbs}, II, D.~G., {Grier}, J.~D.,
  {Hain}, R.~M., {Hall}, D.~M., {Harbo}, P.~N., {(Helen He}, X., {Houck},
  J.~C., {Karovska}, M., {Kashyap}, V.~L., {Lauer}, J., {McCollough}, M.~L.,
  {McDowell}, J.~C., {Miller}, J.~B., {Mitschang}, A.~W., {Morgan}, D.~L.,
  {Mossman}, A.~E., {Nichols}, J.~S., {Nowak}, M.~A., {Plummer}, D.~A.,
  {Refsdal}, B.~L., {Rots}, A.~H., {Siemiginowska}, A., {Sundheim}, B.~A.,
  {Tibbetts}, M.~S., {Van Stone}, D.~W., {Winkelman}, S.~L., \& {Zografou}, P.
  2010, ApJS, 189, 37

\bibitem[{{Geha} {et~al.}(2003){Geha}, {Alcock}, {Allsman}, {Alves}, {Axelrod},
  {Becker}, {Bennett}, {Cook}, {Drake}, {Freeman}, {Griest}, {Keller},
  {Lehner}, {Marshall}, {Minniti}, {Nelson}, {Peterson}, {Popowski}, {Pratt},
  {Quinn}, {Stubbs}, {Sutherland}, {Tomaney}, {Vandehei}, \&
  {Welch}}]{Geha2003AJ}
{Geha}, M., {Alcock}, C., {Allsman}, R.~A., {Alves}, D.~R., {Axelrod}, T.~S.,
  {Becker}, A.~C., {Bennett}, D.~P., {Cook}, K.~H., {Drake}, A.~J., {Freeman},
  K.~C., {Griest}, K., {Keller}, S.~C., {Lehner}, M.~J., {Marshall}, S.~L.,
  {Minniti}, D., {Nelson}, C.~A., {Peterson}, B.~A., {Popowski}, P., {Pratt},
  M.~R., {Quinn}, P.~J., {Stubbs}, C.~W., {Sutherland}, W., {Tomaney}, A.~B.,
  {Vandehei}, T., \& {Welch}, D.~L. 2003, AJ, 125, 1

\bibitem[{{Hawkins}(2002)}]{Hawkins2002MNRAS}
{Hawkins}, M.~R.~S. 2002, MNRAS, 329, 76

\bibitem[{{Heckman} {et~al.}(2004){Heckman}, {Kauffmann}, {Brinchmann},
  {Charlot}, {Tremonti}, \& {White}}]{Heckman2004ApJ}
{Heckman}, T.~M., {Kauffmann}, G., {Brinchmann}, J., {Charlot}, S., {Tremonti},
  C., \& {White}, S.~D.~M. 2004, ApJ, 613, 109

\bibitem[{{Hook} {et~al.}(1994){Hook}, {McMahon}, {Boyle}, \&
  {Irwin}}]{Hook1994MNRAS}
{Hook}, I.~M., {McMahon}, R.~G., {Boyle}, B.~J., \& {Irwin}, M.~J. 1994, MNRAS,
  268, 305

\bibitem[{Hsu {et~al.}(2003)Hsu, Chang, \& Lin}]{libsvmguide}
Hsu, C.-W., Chang, C.-C., \& Lin, C.-J. 2003, A practical guide to support
  vector classification, Tech. rep., Department of Computer Science, National
  Taiwan University

\bibitem[{{Ivezic} {et~al.}(2008){Ivezic}, {Tyson}, {Allsman}, {Andrew},
  {Angel}, \& {for the LSST Collaboration}}]{Ivezic2008arXiv}
{Ivezic}, Z., {Tyson}, J.~A., {Allsman}, R., {Andrew}, J., {Angel}, R., \& {for
  the LSST Collaboration}. 2008, ArXiv e-prints

\bibitem[{{Kaiser}(2004)}]{Kaiser2004SPIE}
{Kaiser}, N. 2004, in Society of Photo-Optical Instrumentation Engineers (SPIE)
  Conference Series, Vol. 5489, Society of Photo-Optical Instrumentation
  Engineers (SPIE) Conference Series, ed. {J.~M.~Oschmann Jr.}, 11--22

\bibitem[{{Kawaguchi} {et~al.}(1998){Kawaguchi}, {Mineshige}, {Umemura}, \&
  {Turner}}]{Kawaguchi1998ApJ}
{Kawaguchi}, T., {Mineshige}, S., {Umemura}, M., \& {Turner}, E.~L. 1998, ApJ,
  504, 671

\bibitem[{{Kelly} {et~al.}(2009){Kelly}, {Bechtold}, \&
  {Siemiginowska}}]{Kelly2009ApJ}
{Kelly}, B.~C., {Bechtold}, J., \& {Siemiginowska}, A. 2009, ApJ, 698, 895

\bibitem[{{Kim} {et~al.}(2011){Kim}, {Protopapas}, {Byun}, {Alcock}, {Khardon},
  \& {Trichas}}]{Kim2011ApJ}
{Kim}, D.-W., {Protopapas}, P., {Byun}, Y.-I., {Alcock}, C., {Khardon}, R., \&
  {Trichas}, M. 2011, ApJ, 735, 68

\bibitem[{{Kollmeier} {et~al.}(2006){Kollmeier}, {Onken}, {Kochanek}, {Gould},
  {Weinberg}, {Dietrich}, {Cool}, {Dey}, {Eisenstein}, {Jannuzi}, {Le Floc'h},
  \& {Stern}}]{Kollmeier2006ApJ}
{Kollmeier}, J.~A., {Onken}, C.~A., {Kochanek}, C.~S., {Gould}, A., {Weinberg},
  D.~H., {Dietrich}, M., {Cool}, R., {Dey}, A., {Eisenstein}, D.~J., {Jannuzi},
  B.~T., {Le Floc'h}, E., \& {Stern}, D. 2006, ApJ, 648, 128

\bibitem[{{Koz{\l}owski} \& {Kochanek}(2009)}]{Kozlowski2009ApJ}
{Koz{\l}owski}, S., \& {Kochanek}, C.~S. 2009, ApJ, 701, 508

\bibitem[{{Koz{\l}owski} {et~al.}(2011){Koz{\l}owski}, {Kochanek}, {Jacyszyn},
  {Udalski}, {Szymanski}, {Poleski}, {Kubiak}, {Soszynski}, {Pietrzynski},
  {Wyrzykowski}, {Ulaczyk}, \& {Pietrukowicz}}]{Kozlowski2011arXiv}
{Koz{\l}owski}, S., {Kochanek}, C.~S., {Jacyszyn}, A.~M., {Udalski}, A.,
  {Szymanski}, M.~K., {Poleski}, R., {Kubiak}, M., {Soszynski}, I.,
  {Pietrzynski}, G., {Wyrzykowski}, L., {Ulaczyk}, K., \& {Pietrukowicz}, P.
  2011, ArXiv e-prints:1106.3110

\bibitem[{{Koz{\l}owski} {et~al.}(2010){Koz{\l}owski}, {Kochanek}, {Udalski},
  {Wyrzykowski}, {Soszy{\'n}ski}, {Szyma{\'n}ski}, {Kubiak}, {Pietrzy{\'n}ski},
  {Szewczyk}, {Ulaczyk}, {Poleski}, \& {The OGLE
  Collaboration}}]{Kozlowski2010ApJ}
{Koz{\l}owski}, S., {Kochanek}, C.~S., {Udalski}, A., {Wyrzykowski}, {\L}.,
  {Soszy{\'n}ski}, I., {Szyma{\'n}ski}, M.~K., {Kubiak}, M., {Pietrzy{\'n}ski},
  G., {Szewczyk}, O., {Ulaczyk}, K., {Poleski}, R., \& {The OGLE
  Collaboration}. 2010, ApJ, 708, 927

\bibitem[{{Lacy} {et~al.}(2004){Lacy}, {Storrie-Lombardi}, {Sajina},
  {Appleton}, {Armus}, {Chapman}, {Choi}, {Fadda}, {Fang}, {Frayer},
  {Heinrichsen}, {Helou}, {Im}, {Marleau}, {Masci}, {Shupe}, {Soifer},
  {Surace}, {Teplitz}, {Wilson}, \& {Yan}}]{Lacy2004ApJS}
{Lacy}, M., {Storrie-Lombardi}, L.~J., {Sajina}, A., {Appleton}, P.~N.,
  {Armus}, L., {Chapman}, S.~C., {Choi}, P.~I., {Fadda}, D., {Fang}, F.,
  {Frayer}, D.~T., {Heinrichsen}, I., {Helou}, G., {Im}, M., {Marleau}, F.~R.,
  {Masci}, F., {Shupe}, D.~L., {Soifer}, B.~T., {Surace}, J., {Teplitz}, H.~I.,
  {Wilson}, G., \& {Yan}, L. 2004, ApJS, 154, 166

\bibitem[{{Laurent} {et~al.}(2000){Laurent}, {Mirabel}, {Charmandaris},
  {Gallais}, {Madden}, {Sauvage}, {Vigroux}, \& {Cesarsky}}]{Laurent2000AA}
{Laurent}, O., {Mirabel}, I.~F., {Charmandaris}, V., {Gallais}, P., {Madden},
  S.~C., {Sauvage}, M., {Vigroux}, L., \& {Cesarsky}, C. 2000, A\&A, 359, 887

\bibitem[{{MacLeod} {et~al.}(2010){MacLeod}, {Ivezi{\'c}}, {Kochanek},
  {Koz{\l}owski}, {Kelly}, {Bullock}, {Kimball}, {Sesar}, {Westman}, {Brooks},
  {Gibson}, {Becker}, \& {de Vries}}]{MacLeod2010ApJ}
{MacLeod}, C.~L., {Ivezi{\'c}}, {\v Z}., {Kochanek}, C.~S., {Koz{\l}owski}, S.,
  {Kelly}, B., {Bullock}, E., {Kimball}, A., {Sesar}, B., {Westman}, D.,
  {Brooks}, K., {Gibson}, R., {Becker}, A.~C., \& {de Vries}, W.~H. 2010, ApJ,
  721, 1014

\bibitem[{Manevitz \& Yousef(2002)}]{Manevitz2002}
Manevitz, L.~M., \& Yousef, M. 2002, J. Mach. Learn. Res., 2, 139

\bibitem[{{Meixner} {et~al.}(2006){Meixner}, {Gordon}, {Indebetouw}, {Hora},
  {Whitney}, {Blum}, {Reach}, {Bernard}, {Meade}, {Babler}, {Engelbracht},
  {For}, {Misselt}, {Vijh}, {Leitherer}, {Cohen}, {Churchwell}, {Boulanger},
  {Frogel}, {Fukui}, {Gallagher}, {Gorjian}, {Harris}, {Kelly}, {Kawamura},
  {Kim}, {Latter}, {Madden}, {Markwick-Kemper}, {Mizuno}, {Mizuno}, {Mould},
  {Nota}, {Oey}, {Olsen}, {Onishi}, {Paladini}, {Panagia}, {Perez-Gonzalez},
  {Shibai}, {Sato}, {Smith}, {Staveley-Smith}, {Tielens}, {Ueta}, {van Dyk},
  {Volk}, {Werner}, \& {Zaritsky}}]{Meixner2006AJ}
{Meixner}, M., {Gordon}, K.~D., {Indebetouw}, R., {Hora}, J.~L., {Whitney}, B.,
  {Blum}, R., {Reach}, W., {Bernard}, J.-P., {Meade}, M., {Babler}, B.,
  {Engelbracht}, C.~W., {For}, B.-Q., {Misselt}, K., {Vijh}, U., {Leitherer},
  C., {Cohen}, M., {Churchwell}, E.~B., {Boulanger}, F., {Frogel}, J.~A.,
  {Fukui}, Y., {Gallagher}, J., {Gorjian}, V., {Harris}, J., {Kelly}, D.,
  {Kawamura}, A., {Kim}, S., {Latter}, W.~B., {Madden}, S., {Markwick-Kemper},
  C., {Mizuno}, A., {Mizuno}, N., {Mould}, J., {Nota}, A., {Oey}, M.~S.,
  {Olsen}, K., {Onishi}, T., {Paladini}, R., {Panagia}, N., {Perez-Gonzalez},
  P., {Shibai}, H., {Sato}, S., {Smith}, L., {Staveley-Smith}, L., {Tielens},
  A.~G.~G.~M., {Ueta}, T., {van Dyk}, S., {Volk}, K., {Werner}, M., \&
  {Zaritsky}, D. 2006, AJ, 132, 2268

\bibitem[{{Miranda} \& {Macci{\`o}}(2007)}]{Miranda2007MNRAS}
{Miranda}, M., \& {Macci{\`o}}, A.~V. 2007, MNRAS, 382, 1225

\bibitem[{{Persic} {et~al.}(2004){Persic}, {Rephaeli}, {Braito}, {Cappi},
  {Della Ceca}, {Franceschini}, \& {Gruber}}]{Persic2004AA}
{Persic}, M., {Rephaeli}, Y., {Braito}, V., {Cappi}, M., {Della Ceca}, R.,
  {Franceschini}, A., \& {Gruber}, D.~E. 2004, A\&A, 419, 849

\bibitem[{{Platt}(1999)}]{Platt1999}
{Platt}, J.~C. 1999, in Advances in Large Margin Classifiers (MIT Press),
  61--74

\bibitem[{{Rees}(1984)}]{Rees1984ARAA}
{Rees}, M.~J. 1984, ARA\&A, 22, 471

\bibitem[{{Ross} {et~al.}(2009){Ross}, {Shen}, {Strauss}, {Vanden Berk},
  {Connolly}, {Richards}, {Schneider}, {Weinberg}, {Hall}, {Bahcall}, \&
  {Brunner}}]{Ross2009ApJ}
{Ross}, N.~P., {Shen}, Y., {Strauss}, M.~A., {Vanden Berk}, D.~E., {Connolly},
  A.~J., {Richards}, G.~T., {Schneider}, D.~P., {Weinberg}, D.~H., {Hall},
  P.~B., {Bahcall}, N.~A., \& {Brunner}, R.~J. 2009, ApJ, 697, 1634

\bibitem[{{Rowan-Robinson} {et~al.}(2008){Rowan-Robinson}, {Babbedge},
  {Oliver}, {Trichas}, {Berta}, {Lonsdale}, {Smith}, {Shupe}, {Surace},
  {Arnouts}, {Ilbert}, {Le F{\'e}vre}, {Afonso-Luis}, {Perez-Fournon},
  {Hatziminaoglou}, {Polletta}, {Farrah}, \& {Vaccari}}]{Rowan2008MNRAS}
{Rowan-Robinson}, M., {Babbedge}, T., {Oliver}, S., {Trichas}, M., {Berta}, S.,
  {Lonsdale}, C., {Smith}, G., {Shupe}, D., {Surace}, J., {Arnouts}, S.,
  {Ilbert}, O., {Le F{\'e}vre}, O., {Afonso-Luis}, A., {Perez-Fournon}, I.,
  {Hatziminaoglou}, E., {Polletta}, M., {Farrah}, D., \& {Vaccari}, M. 2008,
  MNRAS, 386, 697

\bibitem[{{Rowan-Robinson} {et~al.}(2005){Rowan-Robinson}, {Babbedge},
  {Surace}, {Shupe}, {Fang}, {Lonsdale}, {Smith}, {Polletta}, {Siana},
  {Gonzalez-Solares}, {Xu}, {Owen}, {Davoodi}, {Dole}, {Domingue},
  {Efstathiou}, {Farrah}, {Fox}, {Franceschini}, {Frayer}, {Hatziminaoglou},
  {Masci}, {Morrison}, {Nandra}, {Oliver}, {Onyett}, {Padgett},
  {Perez-Fournon}, {Serjeant}, {Stacey}, \& {Vaccari}}]{Rowan2005AJ}
{Rowan-Robinson}, M., {Babbedge}, T., {Surace}, J., {Shupe}, D., {Fang}, F.,
  {Lonsdale}, C., {Smith}, G., {Polletta}, M., {Siana}, B., {Gonzalez-Solares},
  E., {Xu}, K., {Owen}, F., {Davoodi}, P., {Dole}, H., {Domingue}, D.,
  {Efstathiou}, A., {Farrah}, D., {Fox}, M., {Franceschini}, A., {Frayer}, D.,
  {Hatziminaoglou}, E., {Masci}, F., {Morrison}, G., {Nandra}, K., {Oliver},
  S., {Onyett}, N., {Padgett}, D., {Perez-Fournon}, I., {Serjeant}, S.,
  {Stacey}, G., \& {Vaccari}, M. 2005, AJ, 129, 1183

\bibitem[{{Schmidtke} {et~al.}(1999){Schmidtke}, {Cowley}, {Crane}, {Taylor},
  {McGrath}, {Hutchings}, \& {Crampton}}]{Schmidtke1999AJ}
{Schmidtke}, P.~C., {Cowley}, A.~P., {Crane}, J.~D., {Taylor}, V.~A.,
  {McGrath}, T.~K., {Hutchings}, J.~B., \& {Crampton}, D. 1999, AJ, 117, 927

\bibitem[{Sch\"{o}lkopf {et~al.}(2001)Sch\"{o}lkopf, Platt, Shawe-Taylor,
  Smola, \& Williamson}]{Scholkopf2001}
Sch\"{o}lkopf, B., Platt, J.~C., Shawe-Taylor, J.~C., Smola, A.~J., \&
  Williamson, R.~C. 2001, Neural Comput., 13, 1443

\bibitem[{{Sharp} {et~al.}(2006){Sharp}, {Saunders}, {Smith}, {Churilov},
  {Correll}, {Dawson}, {Farrel}, {Frost}, {Haynes}, {Heald}, {Lankshear},
  {Mayfield}, {Waller}, \& {Whittard}}]{Sharp2006SPIE}
{Sharp}, R., {Saunders}, W., {Smith}, G., {Churilov}, V., {Correll}, D.,
  {Dawson}, J., {Farrel}, T., {Frost}, G., {Haynes}, R., {Heald}, R.,
  {Lankshear}, A., {Mayfield}, D., {Waller}, L., \& {Whittard}, D. 2006, in
  Society of Photo-Optical Instrumentation Engineers (SPIE) Conference Series,
  Vol. 6269, Society of Photo-Optical Instrumentation Engineers (SPIE)
  Conference Series

\bibitem[{{Skrutskie} {et~al.}(2006){Skrutskie}, {Cutri}, {Stiening},
  {Weinberg}, {Schneider}, {Carpenter}, {Beichman}, {Capps}, {Chester},
  {Elias}, {Huchra}, {Liebert}, {Lonsdale}, {Monet}, {Price}, {Seitzer},
  {Jarrett}, {Kirkpatrick}, {Gizis}, {Howard}, {Evans}, {Fowler}, {Fullmer},
  {Hurt}, {Light}, {Kopan}, {Marsh}, {McCallon}, {Tam}, {Van Dyk}, \&
  {Wheelock}}]{Skrutskie2006AJ}
{Skrutskie}, M.~F., {Cutri}, R.~M., {Stiening}, R., {Weinberg}, M.~D.,
  {Schneider}, S., {Carpenter}, J.~M., {Beichman}, C., {Capps}, R., {Chester},
  T., {Elias}, J., {Huchra}, J., {Liebert}, J., {Lonsdale}, C., {Monet}, D.~G.,
  {Price}, S., {Seitzer}, P., {Jarrett}, T., {Kirkpatrick}, J.~D., {Gizis},
  J.~E., {Howard}, E., {Evans}, T., {Fowler}, J., {Fullmer}, L., {Hurt}, R.,
  {Light}, R., {Kopan}, E.~L., {Marsh}, K.~A., {McCallon}, H.~L., {Tam}, R.,
  {Van Dyk}, S., \& {Wheelock}, S. 2006, AJ, 131, 1163

\bibitem[{{Stern} {et~al.}(2005){Stern}, {Eisenhardt}, {Gorjian}, {Kochanek},
  {Caldwell}, {Eisenstein}, {Brodwin}, {Brown}, {Cool}, {Dey}, {Green},
  {Jannuzi}, {Murray}, {Pahre}, \& {Willner}}]{Stern2005ApJ}
{Stern}, D., {Eisenhardt}, P., {Gorjian}, V., {Kochanek}, C.~S., {Caldwell},
  N., {Eisenstein}, D., {Brodwin}, M., {Brown}, M.~J.~I., {Cool}, R., {Dey},
  A., {Green}, P., {Jannuzi}, B.~T., {Murray}, S.~S., {Pahre}, M.~A., \&
  {Willner}, S.~P. 2005, ApJ, 631, 163

\bibitem[{{Trichas} {et~al.}(2009){Trichas}, {Georgakakis}, {Rowan-Robinson},
  {Nandra}, {Clements}, \& {Vaccari}}]{Trichas2009MNRAS}
{Trichas}, M., {Georgakakis}, A., {Rowan-Robinson}, M., {Nandra}, K.,
  {Clements}, D., \& {Vaccari}, M. 2009, MNRAS, 399, 663

\bibitem[{{Trichas} {et~al.}(2010){Trichas}, {Rowan-Robinson}, {Georgakakis},
  {Valtchanov}, {Nandra}, {Farrah}, {Morrison}, {Clements}, \&
  {Waddington}}]{Trichas2010MNRAS}
{Trichas}, M., {Rowan-Robinson}, M., {Georgakakis}, A., {Valtchanov}, I.,
  {Nandra}, K., {Farrah}, D., {Morrison}, G., {Clements}, D., \& {Waddington},
  I. 2010, MNRAS, 405, 2243

\bibitem[{{Udalski} {et~al.}(2008){Udalski}, {Szymanski}, {Soszynski}, \&
  {Poleski}}]{Udalski2008AcA}
{Udalski}, A., {Szymanski}, M.~K., {Soszynski}, I., \& {Poleski}, R. 2008, Acta
  Astronomica, 58, 69

\bibitem[{{Watson} {et~al.}(2009){Watson}, {Schr{\"o}der}, {Fyfe}, {Page},
  {Lamer}, {Mateos}, {Pye}, {Sakano}, {Rosen}, {Ballet}, {Barcons}, {Barret},
  {Boller}, {Brunner}, {Brusa}, {Caccianiga}, {Carrera}, {Ceballos}, {Della
  Ceca}, {Denby}, {Denkinson}, {Dupuy}, {Farrell}, {Fraschetti}, {Freyberg},
  {Guillout}, {Hambaryan}, {Maccacaro}, {Mathiesen}, {McMahon}, {Michel},
  {Motch}, {Osborne}, {Page}, {Pakull}, {Pietsch}, {Saxton}, {Schwope},
  {Severgnini}, {Simpson}, {Sironi}, {Stewart}, {Stewart}, {Stobbart}, {Tedds},
  {Warwick}, {Webb}, {West}, {Worrall}, \& {Yuan}}]{Watson2009AA}
{Watson}, M.~G., {Schr{\"o}der}, A.~C., {Fyfe}, D., {Page}, C.~G., {Lamer}, G.,
  {Mateos}, S., {Pye}, J., {Sakano}, M., {Rosen}, S., {Ballet}, J., {Barcons},
  X., {Barret}, D., {Boller}, T., {Brunner}, H., {Brusa}, M., {Caccianiga}, A.,
  {Carrera}, F.~J., {Ceballos}, M., {Della Ceca}, R., {Denby}, M., {Denkinson},
  G., {Dupuy}, S., {Farrell}, S., {Fraschetti}, F., {Freyberg}, M.~J.,
  {Guillout}, P., {Hambaryan}, V., {Maccacaro}, T., {Mathiesen}, B., {McMahon},
  R., {Michel}, L., {Motch}, C., {Osborne}, J.~P., {Page}, M., {Pakull}, M.~W.,
  {Pietsch}, W., {Saxton}, R., {Schwope}, A., {Severgnini}, P., {Simpson}, M.,
  {Sironi}, G., {Stewart}, G., {Stewart}, I.~M., {Stobbart}, A.-M., {Tedds},
  J., {Warwick}, R., {Webb}, N., {West}, R., {Worrall}, D., \& {Yuan}, W. 2009,
  A\&A, 493, 339

\bibitem[{{Zaritsky} {et~al.}(2004){Zaritsky}, {Harris}, {Thompson}, \&
  {Grebel}}]{Zaritsky2004AJ}
{Zaritsky}, D., {Harris}, J., {Thompson}, I.~B., \& {Grebel}, E.~K. 2004, AJ,
  128, 1606

\end{thebibliography}

\end{document}